\documentclass[pre,twocolumn]{revtex4}
\usepackage{amssymb}
\usepackage{amsfonts}
\usepackage{amsmath}
\usepackage{graphicx}
\usepackage{color}

\begin{document}

\title{Hybrid quantum-classical dynamics with stationary thermal states}
\date{\today }
\author{Adri\'{a}n A. Budini}
\affiliation{Consejo Nacional de Investigaciones Cient\'{\i}ficas y T\'{e}cnicas
(CONICET), Centro At\'{o}mico Bariloche, Avenida E. Bustillo Km 9.5, (8400)
Bariloche, Argentina, and Universidad Tecnol\'{o}gica Nacional (UTN-FRC),
Fanny Newbery 111, (8400) Bariloche, Argentina}

\begin{abstract}
Quantum and classical systems can consistently be coupled via non-unitary
time-irreversible mechanisms. In this paper we characterize which kind of
corresponding dynamics converge in the stationary regime to a thermal hybrid
state, that is, a density matrix that maximizes the hybrid arrangement
entropy under the constraints of a canonical ensemble. Introducing a
detailed balance condition, it is found that a specific subclass of hybrid
Lindblad equations fulfill the demanded requirement. The main theoretical
results are exemplified through a set of specific dynamics that in addition
enlighten how the thermal state of each subsystem in isolation is affected
by their mutual coupling. In particular, a Gaussian thermal state could
become a bimodal distribution when increasing the interaction strength of a
classical subsystem with a quantum two-level subsystem.
\end{abstract}

\maketitle

\section{Introduction}

Hybrid quantum-classical dynamics describe the interaction between a
subsystem that has associated a Hilbert space, which implies that its
dynamics is intrinsically open and quantum~\cite{breuerbook}, and a
subsystem that can be completely characterized in terms of probabilities~%
\cite{kampen}. Their coupling can consistently be formulated when taking
into account time-irreversible processes. Most of the advancements in the
characterization of these dynamics rely on two complementary assumptions
about the nature (discrete vs. continuous) of the classical (incoherent)
subsystem.

For discrete classical degrees of freedom the most general time-evolution of
the hybrid state was determined in Ref.~\cite{rate} and subsequently in~\cite%
{breuerRate}. A similar class of hybrid Lindblad equations was also
presented in Ref.~\cite{sudarshan}. Along time, these kind of dynamics
emerged in different physical contexts and situations such as for example in
measurement theory~\cite{measurement}, in the formulation of Bloch-Boltzmann
equations~\cite{alickipaper}, non-Markovian master equations~\cite{vega}
induced by complex structured reservoirs~\cite%
{esposito,random,breuerPreSpin,spinbaths}, single-molecule spectroscopy~\cite%
{barkaiChem,sms,smsJumps}, quantum state smoothing~\cite%
{smoothWise,smoothRate}, as well as in the thermodynamics induced by finite
baths~\cite{sanpera,pekola}.

Complementarily, when the state of classical subsystem can be labeled by a
continuous coordinate the most general possible hybrid time-evolution (a
quantum Fokker-Planck equation) was established recently in Refs.~\cite%
{oppen1,oppen2,lastDiosi,soda}. This case emerges naturally when studying
different physical problems such as the interaction between quantum matter
and classical gravitational fields~\cite%
{karol,penrose,LajosConfer,bassi,tilloyDiosi,oppenGravity1,oppenGravity2,oppenX,JL3}%
, continuous-in-time measurement processes~\cite{halliwel,strunz,diosi}, and
more specifically measurements with continuous feedback control~\cite{QFP}.
In this last context, under some general conditions~\cite{landi}, a recorded
measurement signal can be treated as a classical incoherent system (discrete
or continuous) coupled to the monitored quantum system~\cite%
{landihybrid,landihybridBIS}.

Given their intrinsic theoretical and practical interest, the study of
hybrid quantum-classical dynamics remains as a topic of active research. For
example, the characterization of all possible coupling mechanisms and their
diffusive limit were performed in Ref.~\cite{budini}. Gleason and Kraus
theorems were extended to the hybrid case in Ref.~\cite{camelet}, while
control theory of hybrid systems in~\cite{control}. The study of the entropy
and thermodynamics ensembles associated to hybrid arrangements was initiated
in Refs.~\cite{JL,JL0,tronci,layton,JL5}.

Following the last line of fundamental research, the goal of the present
contribution is to provide a complete and closed description of thermal
hybrid states. Similarly to the standard and well established mechanical
statistical formalism~\cite{reichl} here we define a hybrid state as thermal
if it maximizes entropy under the constraints of a canonical ensemble. Of
special interest is to study how the thermal state of each subsystem in
isolation is affected by the interaction with the other one. Furthermore, as
a main result, we find and characterize time-evolutions that in the
long-time regime converge to a hybrid thermal state.

The main theoretical tool that allows the present results to be obtained in
an embedding of hybrid states in a bipartite Hilbert space. Hence, the
classical system is represented by a quantum subsystem that in a fixed basis
never develops any coherence, that is, its partial state is always diagonal.
Equivalently, the classical subsystem is a quantum one that is always
incoherent in a fixed basis.

Using standard tools of open quantum system theory the\ previous frame
allows us to embed the dynamics in a (bipartite) Lindblad evolution~\cite%
{breuerbook} that fulfill the desired properties. We show that there exist
alternative dynamics and coupling mechanisms that lead to a thermal
stationary state. The main ingredient is a detailed balance condition~\cite%
{breuerbook,alicki} on the possible transitions between both subsystems.
Relevant examples explicitly show the validity and applicability of the
present analysis.

The manuscript is outlined as follows. In Sec. II we introduce the hybrid
quantum-classical states and their associated entropy. In Sec. III,
consistently with a canonical ensemble, thermal hybrid states are explicitly
defined and constructed. In Sec. IV we derive which evolutions in a
stationary regime reach a hybrid thermal state. In Sec. V we study some
examples that enlighten the main theoretical results. Special interest is
paid to a classical harmonic system coupled to a quantum two-level system.
The conclusions are provided in Sec. VI. Extra supporting results are
provided in the Appendixes.

\section{Entropy of hybrid quantum-classical states}

We model a hybrid quantum-classical arrangement through a bipartite quantum
representation consisting of two parts, the quantum subsystem ($\mathcal{S}$%
) and extra degrees of freedom that are associated to the classical
subsystem\ ($\mathcal{C}$). Consistently, the total Hilbert $\mathcal{H}_{%
\mathcal{SC}}$ is the product of each subsystem Hilbert space $\mathcal{H}_{%
\mathcal{SC}}=\mathcal{H}_{\mathcal{S}}\otimes \mathcal{H}_{\mathcal{C}}.$ A
bipartite density matrix $\Xi $ represents a hybrid quantum-classical state
when it assumes the separable form%
\begin{equation}
\Xi =\sum_{c}\rho ^{c}\otimes |c\rangle \langle c|.  \label{BipartiteState}
\end{equation}%
Here, $\{|c\rangle \}$ is an (fixed) orthogonal basis in $\mathcal{H}_{C},$ $%
\langle c|c^{\prime }\rangle =\delta _{cc^{\prime }},$ which in turn
fulfills $\sum_{c}|c\rangle \langle c|=\mathrm{I}_{c},$ where $\mathrm{I}$
denote the identity operator. From now on, we use the index $c$ to label all
possible states of the classical (incoherent) subsystem $\mathcal{C.}$
Furthermore, $\{\rho ^{c}\}$ are conditional quantum states in $\mathcal{H}%
_{S}.$ Notice that each state $\rho ^{c}$ can be read as the (unnormalized)
density matrix of $\mathcal{S}$ \textquotedblleft given
that\textquotedblright\ $\mathcal{C}$ is in the state $|c\rangle \langle c|.$

The normalization of the bipartite state reads%
\begin{equation}
\mathrm{Tr}[\Xi ]=\sum_{c}p^{c}=1,\ \ \ \ \ \ \ \ \ p^{c}\equiv \mathrm{Tr}%
[\rho ^{c}\otimes |c\rangle \langle c|].
\end{equation}%
With $\mathrm{Tr}[\bullet ]$ we denote a trace operation. When necessary, a
subindex in the trace operation denotes a partial trace over the
corresponding subsystem. The partial state of the quantum subsystem is%
\begin{equation}
\rho \equiv \mathrm{Tr}_{\mathcal{C}}[\Xi ]=\sum_{c}\rho ^{c}=\sum_{c}p^{c}%
\bar{\rho}^{c}.
\end{equation}%
Notice that in consequence the state of the quantum subsystem can always be
read as a statistical superposition (mixture) of the normalized states $\bar{%
\rho}^{c}\equiv \rho ^{c}/p_{c}.$ Similarly, the partial state of the
classical subsystem reads%
\begin{equation}
\sigma \equiv \mathrm{Tr}_{\mathcal{S}}[\Xi ]=\sum_{c}p^{c}|c\rangle \langle
c|.
\end{equation}%
Hence, the vector of probabilities $\{p^{c}\}$ gives the statistical
description of the classical subsystem.

Given the bipartite representation, the entropy of a bipartite state $\Xi $
can be defined in a standard way,%
\begin{equation}
S[\Xi ]=-k_{B}\mathrm{Tr}[\Xi \ln (\Xi )],  \label{Entropia}
\end{equation}%
where $k_{B}$ is the Boltzmann constant. For the state~(\ref{BipartiteState}%
), after performing the trace over the classical degrees of freedom, it
follows%
\begin{equation}
S[\Xi ]=-k_{B}\sum_{c}\mathrm{Tr}[\rho ^{c}\ln (\rho ^{c})].
\label{BipartiteEntropy}
\end{equation}%
In general, the entropy depends on which correlations are established
between both subsystems. In fact, in terms of the normalized quantum states $%
\bar{\rho}^{c}=\rho ^{c}/p_{c}$ the entropy reads%
\begin{equation}
S[\Xi ]=-k_{B}\sum_{c}p^{c}\mathrm{Tr}[\bar{\rho}^{c}\ln (\bar{\rho}%
^{c})]-k_{B}\sum_{c}p^{c}\ln (p^{c}).
\end{equation}%
Only in the uncorrelated case%
\begin{equation}
\Xi =\rho \otimes \sum_{c}p^{c}|c\rangle \langle c|,
\end{equation}%
the entropy~(\ref{BipartiteEntropy}) becomes the addition of the entropy of
each subsystem%
\begin{equation}
S[\Xi ]=-k_{B}\mathrm{Tr}[\rho \ln (\rho )]-k_{B}\sum_{c}p^{c}\ln (p^{c}).
\end{equation}%
This case is recovered when the unnormalized quantum states\ $\rho ^{c}$ in
Eq.~(\ref{BipartiteState}) can be written as $\rho ^{c}=p^{c}\rho ,$ where $%
\mathrm{Tr}[\rho ]=1.$

\section{Hybrid thermal states}

Thermal states maximize entropy under specific constraints. In the canonical
ensemble, the constraints are $\mathrm{Tr}[\Xi ]=1$ and $\mathrm{Tr}[\Xi 
\mathcal{H}]=\langle \mathcal{H}\rangle ,$ where $\mathcal{H}$ is the
Hamiltonian operator. As is well known~\cite{reichl}, entropy maximization
is achieved by the state%
\begin{equation}
\Xi _{\mathrm{th}}=\frac{e^{-\beta \mathcal{H}}}{\mathrm{Tr}[e^{-\beta 
\mathcal{H}}]},  \label{ThermalArbitrario}
\end{equation}%
where $\beta \equiv 1/k_{B}T.$ As usual, $T$ is the temperature. After
defining the Hamiltonian $\mathcal{H}$, this general result allows us to
define thermal states for hybrid quantum-classical arrangements.

The Hamiltonian operator of the hybrid arrangement is taken as%
\begin{equation}
\mathcal{H}=\sum_{c}E_{c}|c\rangle \langle c|+H_{s}+\lambda \sum_{c}\bar{H}%
_{c}\otimes |c\rangle \langle c|.  \label{HGeneralHybrido}
\end{equation}%
For simplicity, the classical subsystem is taken as a discrete one. In fact, 
$\{E_{c}\}$ define the energies of each state of the classical subsystem,
while $H_{s}$ is the Hamiltonian of the quantum subsystem when isolated. The
dimensionless parameter $\lambda $ measures the unitary coupling between
both subsystems, which has associated the set of arbitrary Hamiltonians
operators $\{\bar{H}_{c}\}.$

We notice that Eq.~(\ref{HGeneralHybrido}) defines a general class of
possible Hamiltonians for the hybrid arrangement, where each subpart has its
own Hamiltonian which in turn is modified by their mutual interaction. The
total Hamiltonian can always be rewritten as%
\begin{equation}
\mathcal{H}=\sum_{c}H_{c}\otimes |c\rangle \langle c|,  \label{Hhibrido}
\end{equation}%
where $H_{c}\equiv E_{c}\mathrm{I}_{s}+H_{s}+\lambda \bar{H}_{c}.$
Introducing this last expression for $\mathcal{H}$ into Eq.~(\ref%
{ThermalArbitrario}), it follows%
\begin{equation}
\Xi _{\mathrm{th}}=\frac{\sum_{c}e^{-\beta H_{c}}\otimes |c\rangle \langle c|%
}{\sum_{c^{\prime }}\mathrm{Tr}[e^{-\beta H_{c^{\prime }}}]}.
\label{ThermalBipa}
\end{equation}%
Notice that the partition function is the addition of the partition
functions associated to the set of Hamiltonians $\{H_{c}\}.$ The hybrid
thermal state $\Xi _{\mathrm{th}}$ can be rewritten as%
\begin{equation}
\Xi _{\mathrm{th}}=\sum_{c}w_{c}\frac{e^{-\beta H_{c}}}{\mathrm{Tr}%
[e^{-\beta H_{c}}]}\otimes |c\rangle \langle c|.  \label{ThermalPesado}
\end{equation}%
Thus, it corresponds to a statistical mixture of the quantum thermal states $%
e^{-\beta H_{c}}/\mathrm{Tr}[^{-\beta H_{c}}]$ associated to each classical
state $|c\rangle \langle c|.$ The weight of each contribution is%
\begin{equation}
w_{c}\equiv \frac{\mathrm{Tr}[e^{-\beta H_{c}}]}{\sum_{c^{\prime }}\mathrm{Tr%
}[e^{-\beta H_{c^{\prime }}}]}.  \label{WCp}
\end{equation}%
Eq.~(\ref{ThermalPesado}) implies that neither the quantum or classical
subsystems are in a thermal state. In fact, the quantum state reads%
\begin{equation}
\rho _{\mathrm{th}}\equiv \mathrm{Tr}_{\mathcal{C}}[\Xi _{\mathrm{th}%
}]=\sum_{c}w_{c}\frac{e^{-\beta H_{c}}}{\mathrm{Tr}_{\mathcal{S}}[e^{-\beta
H_{c}}]},  \label{RhoThS}
\end{equation}%
while the classical state is%
\begin{equation}
\sigma _{\mathrm{th}}\equiv \mathrm{Tr}_{\mathcal{S}}[\Xi _{\mathrm{th}%
}]=\sum_{c}w_{c}|c\rangle \langle c|.  \label{ThermalClasico}
\end{equation}

We notice that in Eq.~(\ref{ThermalPesado}) the conditional thermal states $%
e^{-\beta H_{c}}/\mathrm{Tr}[e^{-\beta H_{c}}]=e^{-\beta (H_{s}+\lambda \bar{%
H}_{c})}/\mathrm{Tr}[e^{-\beta (H_{s}+\lambda \bar{H}_{c})}]$ do not depends
on the energies $\{E_{c}\}$ of the classical subsystem. Furthermore, using
that $H_{c}=E_{c}\mathrm{I}_{s}+H_{s}+\lambda \bar{H}_{c},$ the weights $%
\{w_{c}\}$ [Eq.~(\ref{WCp})] can be rewritten as%
\begin{equation}
w_{c}=\frac{e^{-\beta (E_{c}+F_{c})}}{Z_{\mathrm{th}}}=\frac{e^{-\beta E_{c}}%
}{Z}\left( \frac{e^{-\beta F_{c}}}{Z_{\mathrm{th}}/Z}\right) ,
\label{PesosWc}
\end{equation}%
where $Z_{\mathrm{th}}\equiv \sum_{c}e^{-\beta (E_{c}+F_{c})},$ $Z\equiv
\sum_{c}e^{-\beta E_{c}},$ and $F_{c}$ is the Helmholtz free energy~\cite%
{reichl} associated to each sub-Hamiltonian,%
\begin{equation}
F_{c}\equiv -k_{B}T\ln \mathrm{Tr}[e^{-\beta (H_{s}+\lambda \bar{H}_{c})}].
\end{equation}%
These last two equations give a simple physical frame for understanding the
structure of the bipartite thermal state~(\ref{ThermalPesado}). The quantum
state becomes a statistical mixture of thermal states while the
thermodynamic of the classical system is shifted by the Helmholtz free
energy associated to each quantum sub-Hamiltonian. Eqs.~(\ref{RhoThS}) and~(%
\ref{ThermalClasico}) define the structural changes induced by the
interaction between both subsystems. In fact, only in the non-interacting
case\textit{, }$\lambda =0$\ in Eq.~(\ref{HGeneralHybrido}), the bipartite
thermal state~(\ref{ThermalPesado}) becomes%
\begin{equation}
\Xi _{\mathrm{th}}=\frac{e^{-\beta H_{s}}}{\mathrm{Tr}[e^{-\beta H_{s}}]}%
\otimes \frac{\sum_{c}e^{-\beta E_{c}}|c\rangle \langle c|}{\sum_{c^{\prime
}}e^{-\beta E_{c^{\prime }}}},  \label{ThermalSeparable}
\end{equation}%
that is, the uncorrelated product of each subsystem thermal state.

\section{Dynamics that lead to stationary hybrid thermal states}

In this section we establish which kind of time-irreversible dynamics has
associated an stationary hybrid thermal state [Eq.~(\ref{ThermalBipa}) or
equivalently~(\ref{ThermalPesado})]. In a first step, we review which kind
of dynamics have a stationary thermal state~\cite{breuerbook,alicki}. Our
analysis relies on assuming that both the quantum and classical subsystems
are of \textit{finite dimension}. Hence, the complementary cases (infinite
dimension and/or continuous limits) are defined by taking the appropriate
limits (see for example Sec. V).

\subsection{Density matrix evolution with stationary thermal states \label%
{ThermalSec}}

Given an Hilbert space of finite dimension, and given an arbitrary
Hamiltonian $\mathcal{H}$, there always exist a Lindblad evolution that
leads to a stationary thermal state [Eq.~(\ref{ThermalArbitrario})]. The
eigenvalues and eigenvectors of $\mathcal{H}$ are denoted with with $%
\{\varepsilon _{j}\}$ and $\{|j\rangle \}$ respectively. The evolution of
the time-dependent density matrix $\Xi _{t}$ is written as%
\begin{equation}
\frac{d\Xi _{t}}{dt}=\mathcal{L}_{\mathrm{th}}[\Xi _{t}]=\sum_{\{i,j\}}%
\mathcal{L}_{\{i,j\}}[\Xi _{t}].  \label{ThermalLin}
\end{equation}%
In order to obtain a description closer to classical systems, $\mathcal{L}_{%
\mathrm{th}}$ has been written as an addition over the eigensystem of $%
\mathcal{H.}$ In fact, $\sum_{\{i,j\}}$ is an addition over all possible
pair of eigenstates $\{i,j\}.$ Instead, alternative (equivalent)
presentations rely on an addition over frequencies~\cite{alicki}, $%
\sum_{\{i,j\}}\mathcal{L}_{\{i,j\}}\leftrightarrow \sum_{\omega }\mathcal{L}%
_{\omega },$ where $\omega \leftrightarrow \varepsilon _{i}-\varepsilon
_{j}. $ In this way, introducing the operators $A_{ij}\equiv |i\rangle
\langle j|$ $(i\neq j)$ each contribution in Eq.~(\ref{ThermalLin}) is
written as%
\begin{eqnarray}
\mathcal{L}_{\{i,j\}}[\bullet ] &=&+\gamma _{ij}\left( A_{ij}\bullet
A_{ij}^{\dagger }-\frac{1}{2}\left\{ A_{ij}^{\dagger }A_{ij},\bullet
\right\} _{+}\right)  \notag \\
&&+\gamma _{ji}\left( A_{ij}^{\dagger }\bullet A_{ij}-\frac{1}{2}\left\{
A_{ij}A_{ij}^{\dagger },\bullet \right\} _{+}\right) ,\ \ \ \ 
\label{SuperTh}
\end{eqnarray}%
where $\{a,b\}_{+}\equiv ab+ba$ is the anticommutator operation. The
superoperartor~(\ref{SuperTh}) has the structure of a Lindblad equation. In
fact, $\gamma _{ij}$ defines the rate for the transition $|j\rangle
\rightarrow |i\rangle $ while $\gamma _{ji}$ defines the rate for the
transition $|i\rangle \rightarrow |j\rangle .$ In agreement with a quantum
detailed balance condition~\cite{alicki}, they must satisfy%
\begin{equation}
\frac{\gamma _{ij}}{\gamma _{ji}}=\exp [-\beta (\varepsilon _{i}-\varepsilon
_{j})],  \label{ThermalRates}
\end{equation}%
or equivalently, in a form closer to classical systems~\cite{kampen}, the
detailed balance condition%
\begin{equation}
\gamma _{ij}e^{-\beta \varepsilon _{j}}=\gamma _{ji}e^{-\beta \varepsilon
_{i}}.
\end{equation}

Using that $A_{ij}=|i\rangle \langle j|,$ the superoperator~(\ref{SuperTh})
can explicitly be written as%
\begin{eqnarray}
\mathcal{L}_{\{i,j\}}[\bullet ] &=&+\gamma _{ij}\left( |i\rangle \langle
j|\bullet |j\rangle \langle i|-\frac{1}{2}\left\{ |j\rangle \langle
j|,\bullet \right\} _{+}\right)  \notag \\
&&+\gamma _{ji}\left( |j\rangle \langle i|\bullet |i\rangle \langle j|-\frac{%
1}{2}\left\{ |i\rangle \langle i|,\bullet \right\} _{+}\right) .\ \ \ \ 
\label{Lij}
\end{eqnarray}%
From here, it is simple to check that the thermal state~(\ref%
{ThermalArbitrario}), that is%
\begin{equation}
\langle j|\Xi _{\mathrm{th}}|j\rangle =\frac{e^{-\beta \varepsilon _{j}}}{z}%
,\ \ \ \ \ \langle i|\Xi _{\mathrm{th}}|j\rangle =0,
\end{equation}%
where $z\equiv \sum_{j}e^{-\beta \varepsilon _{j}},$ under the conditions~(%
\ref{ThermalRates}) fulfills $\mathcal{L}_{\mathrm{th}}[\Xi _{\mathrm{th}%
}]=0.$ Hence, it is the stationary solution of Eq.~(\ref{ThermalLin}), $\Xi
_{\mathrm{th}}=\lim_{t\rightarrow \infty }\Xi _{t}.$

It is relevant to remark that Eq.~(\ref{ThermalLin}) could be generalized
(or rewritten) as $\mathcal{L}_{\mathrm{th}}\rightarrow \sum q_{\alpha }%
\mathcal{L}_{\alpha },$ where each $\mathcal{L}_{\alpha }$ is consistent
with detailed balance [Eqs. (\ref{SuperTh}) and (\ref{ThermalRates})] and $%
\{q_{\alpha }\}$ are positive weighs~\cite{alicki}. This property emerges in
the studied examples (Sec. V).

\subsection{Hybrid thermal evolutions}

Given the generality of the previous result, it can also be applied to
hybrid evolutions. We assume the Hamiltonian defined by Eq.~(\ref{Hhibrido}%
), $\mathcal{H}=\sum_{c}H_{c}\otimes |c\rangle \langle c|.$ Its eigensystem
is defined as%
\begin{equation}
\mathcal{H}|j,c\rangle =\varepsilon _{j}^{(c)}|j,c\rangle ,
\end{equation}%
where $\varepsilon _{j}^{(c)}$ is the $j$-eigenvalue of the Hamiltonian $%
H_{c}.$ Thus, under the association $j\rightarrow (j,c),$ and taking into
account the structure defined by Eq.~(\ref{BipartiteState}), it is possible
to establishing evolutions that in the stationary regime converge to an
hybrid thermal state [Eq.~(\ref{ThermalBipa})].

For simplifying the presentation the explicit derivation is performed in
Appendix~\ref{Derivation}. Here, we present the final result. Introducing
the operators%
\begin{equation}
A_{\tilde{j}i}=|\tilde{j}\rangle \langle i|,\ \ \ \ \ \ A_{i\tilde{j}%
}=|i\rangle \langle \tilde{j}|,  \label{Aes}
\end{equation}%
where $\{|i\rangle \}$ and $\{|\tilde{j}\rangle \}$ are the eigenstates of $%
H_{c}$ and $H_{\tilde{c}}$ respectively $(\tilde{c}\neq c),$ a stationary
thermal hybrid state is achieved when the conditional states $\{\rho ^{c}\}$
in Eq.~(\ref{BipartiteState}) evolve as%
\begin{eqnarray}
\frac{d\rho ^{c}}{dt} &=&\mathcal{L}_{\mathrm{th}}^{(c)}[\rho ^{c}]-\frac{1}{%
2}\sum_{\substack{ \tilde{c}  \\ \tilde{c}\neq c}}\sum_{\{i,\tilde{j}%
\}}\gamma _{\tilde{j}i}^{(\tilde{c}c)}\left\{ A_{\tilde{j}i}^{\dagger }A_{%
\tilde{j}i},\rho ^{c}\right\} _{+}  \notag \\
&&+\sum_{\substack{ \tilde{c}  \\ \tilde{c}\neq c}}\sum_{\{i,\tilde{j}%
\}}\gamma _{i\tilde{j}}^{(c\tilde{c})}A_{i\tilde{j}}\rho ^{\tilde{c}}A_{i%
\tilde{j}}^{\dagger }.  \label{ThermalHybrid}
\end{eqnarray}%
This is the main result of this section. Its structure is specified below.
The isolated action of the Lindblad contribution $\mathcal{L}_{\mathrm{th}%
}^{(c)}$ leads to a stationary thermal state associated to $H_{c}.$ It can
be written as%
\begin{equation}
\mathcal{L}_{\mathrm{th}}^{(c)}[\bullet ]=-i[H_{c},\bullet ]+\sum_{\{i,j\}}%
\mathcal{L}_{\{i,j\}}^{(c)}[\bullet ],
\end{equation}%
where $\mathcal{L}_{\{i,j\}}^{(c)}$ are defined by Eq.~(\ref{SuperTh}) [or
equivalently Eq.~(\ref{Lij})] with operators $A_{ij}=|i\rangle \langle j|$
and under the replacement $\gamma _{ij}\rightarrow \gamma _{ij}^{(c)}.$ The
states $|i\rangle $ and $|j\rangle $ are eigenstates of $H_{c}.$
Consistently with Eq.~(\ref{ThermalRates}), the rates fulfill%
\begin{equation}
\frac{\gamma _{ij}^{(c)}}{\gamma _{ji}^{(c)}}=\exp [-\beta (\varepsilon
_{i}^{(c)}-\varepsilon _{j}^{(c)})],  \label{DiagonalThermalRates}
\end{equation}%
where $\varepsilon _{i}^{(c)}$ and $\varepsilon _{j}^{(c)}$ are the
eigenvalues associated to $|i\rangle $ and $|j\rangle $ respectively.
Similarly, in the non-diagonal contributions (proportional to $A_{i\tilde{j}%
}\rho ^{\tilde{c}}A_{i\tilde{j}}^{\dagger })$ the rates must fulfill%
\begin{equation}
\frac{\gamma _{i\tilde{j}}^{(c\tilde{c})}}{\gamma _{\tilde{j}i}^{(\tilde{c}%
c)}}=\exp [-\beta (\varepsilon _{i}^{(c)}-\varepsilon _{\tilde{j}}^{(\tilde{c%
})})],  \label{NonDiagonalThermalRates}
\end{equation}%
where $\{\varepsilon _{i}^{(c)}\}$ and $\{\varepsilon _{\tilde{j}}^{(\tilde{c%
})}\}$ are the eigenvalues associated to the eigenstates $\{|i\rangle \}$
and $\{|\tilde{j}\rangle \}$ of $H_{c}$ and $H_{\tilde{c}}$ respectively $(%
\tilde{c}\neq c).$ The origin of this relation is the same as for the
diagonal rates. In fact, while Eq.~(\ref{DiagonalThermalRates}) guarantees
the achievement of a thermal state of the quantum subsystem in absence of
transitions of the classical one, Eq.~(\ref{NonDiagonalThermalRates})
guarantees the achievement of a hybrid thermal state when involving
transitions in both subsystems. Both kind of rates are associated
respectively to the contributions $e^{-\beta H_{c}}/\mathrm{Tr}[e^{-\beta
H_{c}}]$ and $w_{c}$ in Eq.~(\ref{ThermalPesado}).

We realize that Eq.~(\ref{ThermalHybrid}) is a particular case of Lindblad
rate equation~\cite{rate} whose coupling mechanisms fall in the categories
described in Ref.~\cite{budini}. The term $\mathcal{L}_{\mathrm{th}}^{(c)}$
depends parametrically on the state of the classical subsystem (first
mechanism), while the remaining structure couples intrinsically the
transitions of both subsystems (third mechanism). The structure of this last
contribution is similar to a classical master equation in the sense that
there are \textquotedblleft \textit{gain}\textquotedblright\ and
\textquotedblleft \textit{loss}\textquotedblright\ terms, that is, the terms
proportional to $\rho ^{\tilde{c}}$ $(\tilde{c}\neq c)$ and $\rho ^{c}$
respectively. The gain contributions take into account the transitions $%
|ic\rangle \leftrightarrow |\tilde{j}\tilde{c}\rangle ,$ which modify the
states of both (quantum and classical) subsystems. The loss contributions
take into account the transitions $|ic\rangle \leftrightarrow |jc\rangle ,$
that is, transitions that do not modify the state of the classical
incoherent subsystem.

The second mechanism of Ref.~\cite{budini}, which consists in transitions of
the classical subsystem that occur independently of the state of the quantum
subsystem, are not present in Eq.~(\ref{ThermalHybrid}). In fact, such kind
of transitions are incompatible with the (classical) thermal weights $%
\{w_{c}\}$ [Eq.~(\ref{PesosWc})], which intrinsically depends on the quantum
subsystem. The fourth mechanism, which is defined by transition operators
that are a linear combination of the second and third mechanisms also are
not present due to the same reason.

The previous construction, which relies on the detailed balance condition,
guarantees that the stationary state of the evolution~(\ref{ThermalHybrid})\
is the hybrid thermal state~(\ref{ThermalPesado}). On the other hand, it is
simple to realize that in general it is not possible to write a closed
(local in time) evolution for the partial states of each subsystem. In
Appendix~\ref{embedding} we show that the evolution of the conditional
states $\{\rho ^{c}\}$ [Eq.~(\ref{ThermalHybrid})] can also be written as a
Lindblad evolution for the bipartite state $\Xi $ [Eq.~(\ref{BipartiteState}%
)].

\subsubsection*{Thermal evolution in terms of a unique base of transition
operators}

The general evolution~(\ref{ThermalHybrid}) is written in terms of operators
that induce transitions between the eigenstates of different coupling
Hamiltonians $\{H_{c}\}$ [see Eq.~(\ref{Aes})]. The structure of the
dynamics is enlightened by writing the evolution of $\rho ^{c}$ only in
terms of the eigenbasis of a unique Hamilton, say $H_{c}.$ Let its
eigenbasis be $\{|i\rangle \}$ and express any other basis $\{|\tilde{\imath}%
\rangle \}$ as $|\tilde{\imath}\rangle =\tilde{U}|i\rangle .$ Notice that
the unitary operator $\tilde{U}$ does not depend on the particular states $%
\{|i\rangle \}$ and $\{|\tilde{\imath}\rangle \}.$ It only depends on the
(chosen) Hamiltonian $H_{c}$ and $H_{\tilde{c}}$ (for notational convenience
this dependence on the classical index is not explicitly written). Eq.~(\ref%
{Aes}) can then be written as%
\begin{equation}
A_{\tilde{j}i}=\tilde{U}A_{ji},\ \ \ \ \ \ A_{i\tilde{j}}=A_{ij}\tilde{U}%
^{\dagger }.
\end{equation}%
Therefore, Eq.~(\ref{ThermalHybrid}) becomes%
\begin{eqnarray}
&&\frac{d\rho ^{c}}{dt}\overset{(c)}{=}\mathcal{L}_{\mathrm{th}}^{(c)}[\rho
^{c}]-\frac{1}{2}\sum_{\substack{ \tilde{c}  \\ \tilde{c}\neq c}}\sum_{\{i,%
\tilde{j}\}}\gamma _{\tilde{j}i}^{(\tilde{c}c)}\left\{ A_{ji}^{\dagger
}A_{ji},\rho ^{c}\right\} _{+}  \notag \\
&&+\sum_{\substack{ \tilde{c}  \\ \tilde{c}\neq c}}\sum_{\{i,\tilde{j}%
\}}\gamma _{i\tilde{j}}^{(c\tilde{c})}A_{ij}\tilde{U}^{\dagger }\rho ^{%
\tilde{c}}\tilde{U}A_{ij}^{\dagger },  \label{GenUU}
\end{eqnarray}%
where $\overset{(c)}{=}$ indicates that a particular eigenbasis was chosen.
From this alternative expression it is possible to read the non-diagonal
couplings $(\tilde{c}\neq c)$ as a kind of collisional dynamics~\cite%
{embedding}\ where the transitions $c\rightarrow \tilde{c}$ are endowed with
the unitary transformation $\rho ^{c}\rightarrow \rho ^{\tilde{c}}=\tilde{U}%
\rho ^{c}\tilde{U}^{\dagger }.$ The inverse transitions $\tilde{c}%
\rightarrow c$ involves the inverse unitary transformation. Notice that the
change of basis introduced by $\tilde{U}$ \textquotedblleft
allows\textquotedblright\ to each state $\rho ^{\tilde{c}}$ to reach the
corresponding thermal state associated to each Hamiltonian $H_{\tilde{c}}.$

\section{Examples}

Here, we study different examples that enlighten the main results. For
achieving a deeper understanding of the previous general results first we
consider that both the quantum and classical subsystems are two-dimensional.
After this step, we study a lattice dynamics which allows to describing a
minimal model for a hybrid evolution consisting on a quantum two-level
system interacting with a harmonic classical (incoherent) variable.

\subsection{Dichotomic classical degrees of freedom}

Here we consider a quantum two-level system (qubit) that in turn is coupled
to a dichotomic classical (incoherent) degree of freedom. The hybrid
Hamiltonian [Eq.~(\ref{HGeneralHybrido})]\ reads%
\begin{equation}
\mathcal{H}=\sum_{c=a,b}E_{c}|c\rangle \langle c|+\sum_{c=a,b}\bar{H}%
_{c}\otimes |c\rangle \langle c|.  \label{HDico}
\end{equation}%
The index $c=a,b,$ labels the two states of the classical subsystem. $\bar{H}%
_{a}$ and $\bar{H}_{b}$ are two arbitrary Hamiltonians. For clarifying the
role of the energies of the classical system $\{E_{c}\}$ and those
corresponding to the quantum system (here the eigenvalues of $\{\bar{H}%
_{c}\})$ we have taken $H_{s}=0$ and $\lambda =1$ [see Eq.~(\ref%
{HGeneralHybrido})]. Nevertheless, these conditions can be raised up with a
simple change of notation, $\bar{H}_{c}\rightarrow H_{s}+\lambda \bar{H}%
_{c}. $

The evolution of the conditional\ states $\rho ^{c}$ is taken as $(c\neq 
\tilde{c})$%
\begin{subequations}
\label{GeneralTLS}
\begin{eqnarray}
\frac{d\rho ^{c}}{dt} &=&-i[\bar{H}_{c},\rho ^{c}]+\mathcal{L}_{\mathrm{th}%
}^{(c)}[\rho ^{c}] \\
&&-\frac{1}{2}\gamma _{\downarrow }^{\tilde{c}c}\left\{ \sigma _{c}^{\dagger
}\sigma _{c},\rho ^{c}\right\} _{+}+\gamma _{\uparrow }^{c\tilde{c}}\sigma
_{c}^{\dagger }\tilde{U}^{\dagger }\rho ^{\tilde{c}}\tilde{U}\sigma _{c} \\
&&-\frac{1}{2}\gamma _{\uparrow }^{\tilde{c}c}\left\{ \sigma _{c}\sigma
_{c}^{\dagger },\rho ^{c}\right\} _{+}+\gamma _{\downarrow }^{c\tilde{c}%
}\sigma _{c}\tilde{U}^{\dagger }\rho ^{\tilde{c}}\tilde{U}\sigma
_{c}^{\dagger } \\
&&-\frac{1}{2}\gamma _{+}^{\tilde{c}c}\left\{ \Pi _{c}^{+},\rho ^{c}\right\}
_{+}+\gamma _{+}^{c\tilde{c}}\Pi _{c}^{+}\tilde{U}^{\dagger }\rho ^{\tilde{c}%
}\tilde{U}\Pi _{c}^{+} \\
&&-\frac{1}{2}\gamma _{-}^{\tilde{c}c}\left\{ \Pi _{c}^{-},\rho ^{c}\right\}
_{+}+\gamma _{-}^{c\tilde{c}}\Pi _{c}^{-}\tilde{U}^{\dagger }\rho ^{\tilde{c}%
}\tilde{U}\Pi _{c}^{-}.
\end{eqnarray}%
The lowering operator $\sigma _{c}=|-\rangle _{c}\langle +|_{c},$ raising
operator $\sigma _{c}^{\dagger }=|+\rangle _{c}\langle -|_{c},$ and
projectors $\Pi _{c}^{\pm }=|\pm \rangle _{c}\langle \pm |_{c}$ are defined
in terms of the (two) eigenstates of $\bar{H}_{c},$ that is, $\bar{H}%
_{c}|\pm \rangle _{c}=\varepsilon _{\pm }^{(c)}|\pm \rangle _{c}.$ The
diagonal thermal Lindblad contribution is 
\end{subequations}
\begin{eqnarray}
\mathcal{L}_{\mathrm{th}}^{(c)}[\rho ] &=&\gamma _{\downarrow }^{c}(\sigma
_{c}\rho \sigma _{c}^{\dagger }-\frac{1}{2}\{\sigma _{c}^{\dagger }\sigma
_{c},\rho \}_{+})  \notag \\
&&+\gamma _{\uparrow }^{c}(\sigma _{c}^{\dagger }\rho \sigma _{c}-\frac{1}{2}%
\{\sigma _{c}\sigma _{c}^{\dagger },\rho \}_{+}).
\end{eqnarray}%
The unitary operator $\tilde{U}$ appearing in the non-diagonal contributions
defines the transformation between the two eigenbasis,%
\begin{equation}
\left( 
\begin{array}{c}
|+\rangle _{\tilde{c}} \\ 
|-\rangle _{\tilde{c}}%
\end{array}%
\right) =\tilde{U}\left( 
\begin{array}{c}
|+\rangle _{c} \\ 
|-\rangle _{c}%
\end{array}%
\right) .
\end{equation}%
With this definition it is simple to check that the evolution~(\ref%
{GeneralTLS}) falls into the general structure defined by Eq.~(\ref%
{ThermalHybrid}) [see also Eq.~(\ref{GenUU})]. In fact, this follows after
defining the transition operators $\sigma _{c\tilde{c}}=|-\rangle
_{c}\langle +|_{\tilde{c}}=|-\rangle _{c}\langle +|_{c}\tilde{U}^{\dagger
}=\sigma _{c}\tilde{U}^{\dagger }.$

In Fig.~1 we plot the energy states jointly with all possible transitions
associated to Eq. (\ref{GeneralTLS}). In the first line of this evolution,
the rates $\gamma _{\downarrow }^{c}$\ and $\gamma _{\uparrow }^{c}$\ set
the transitions $|+\rangle _{c}\overset{\gamma _{\downarrow }^{c}}{%
\rightarrow }|-\rangle _{c}$ and $|-\rangle _{c}\overset{\gamma _{\uparrow
}^{c}}{\rightarrow }|+\rangle _{c}$\ respectively [mechanism~(a) in Fig.~1].
The second line introduce the transition $|+\rangle _{c}\overset{\gamma
_{\downarrow }^{\tilde{c}c}}{\rightarrow }|-\rangle _{\tilde{c}}$ and $%
|-\rangle _{\tilde{c}}\overset{\gamma _{\uparrow }^{c\tilde{c}}}{\rightarrow 
}|+\rangle _{c},$ [mechanism~(b) in Fig.~1],\ while the third line
corresponds to $|-\rangle _{c}\overset{\gamma _{\uparrow }^{\tilde{c}c}}{%
\rightarrow }|+\rangle _{\tilde{c}}$ and $|+\rangle _{\tilde{c}}\overset{%
\gamma _{\downarrow }^{c\tilde{c}}}{\rightarrow }|-\rangle _{c}$
[mechanism~(c) in Fig.~1]. The fourth line has associated the transitions $%
|+\rangle _{c}\overset{\gamma _{+}^{\tilde{c}c}}{\rightarrow }|+\rangle _{%
\tilde{c}}$ and $|+\rangle _{\tilde{c}}\overset{\gamma _{+}^{c\tilde{c}}}{%
\rightarrow }|+\rangle _{c}$ [mechanism~(d) in Fig.~1]. Finally, the fifth
line, $|-\rangle _{c}\overset{\gamma _{-}^{\tilde{c}c}}{\rightarrow }%
|-\rangle _{\tilde{c}}$ and $|-\rangle _{\tilde{c}}\overset{\gamma _{-}^{c%
\tilde{c}}}{\rightarrow }|-\rangle _{c}$\ [mechanism~(e) in Fig.~1].
Consistently with Eqs.~(\ref{DiagonalThermalRates}) and~(\ref%
{NonDiagonalThermalRates}), the rates must fulfill 
\begin{subequations}
\label{DetBalTLS}
\begin{eqnarray}
\frac{\gamma _{\uparrow }^{c}}{\gamma _{\downarrow }^{c}} &=&e^{-\beta
\lbrack \varepsilon _{+}^{(c)}-\varepsilon _{-}^{(c)}]}, \\
\frac{\gamma _{\uparrow }^{c\tilde{c}}}{\gamma _{\downarrow }^{\tilde{c}c}}
&=&e^{-\beta \lbrack (E_{c}+\varepsilon _{+}^{(c)})-(E_{\tilde{c}%
}+\varepsilon _{-}^{(\tilde{c})})]}, \\
\frac{\gamma _{\pm }^{c\tilde{c}}}{\gamma _{\pm }^{\tilde{c}c}} &=&e^{-\beta
\lbrack (E_{c}+\varepsilon _{\pm }^{(c)})-(E_{\tilde{c}}+\varepsilon _{\pm
}^{(\tilde{c})})]}.
\end{eqnarray}
\begin{figure}[tbp]
\includegraphics[bb=0 25 495 320,
angle=0,width=8.5cm]{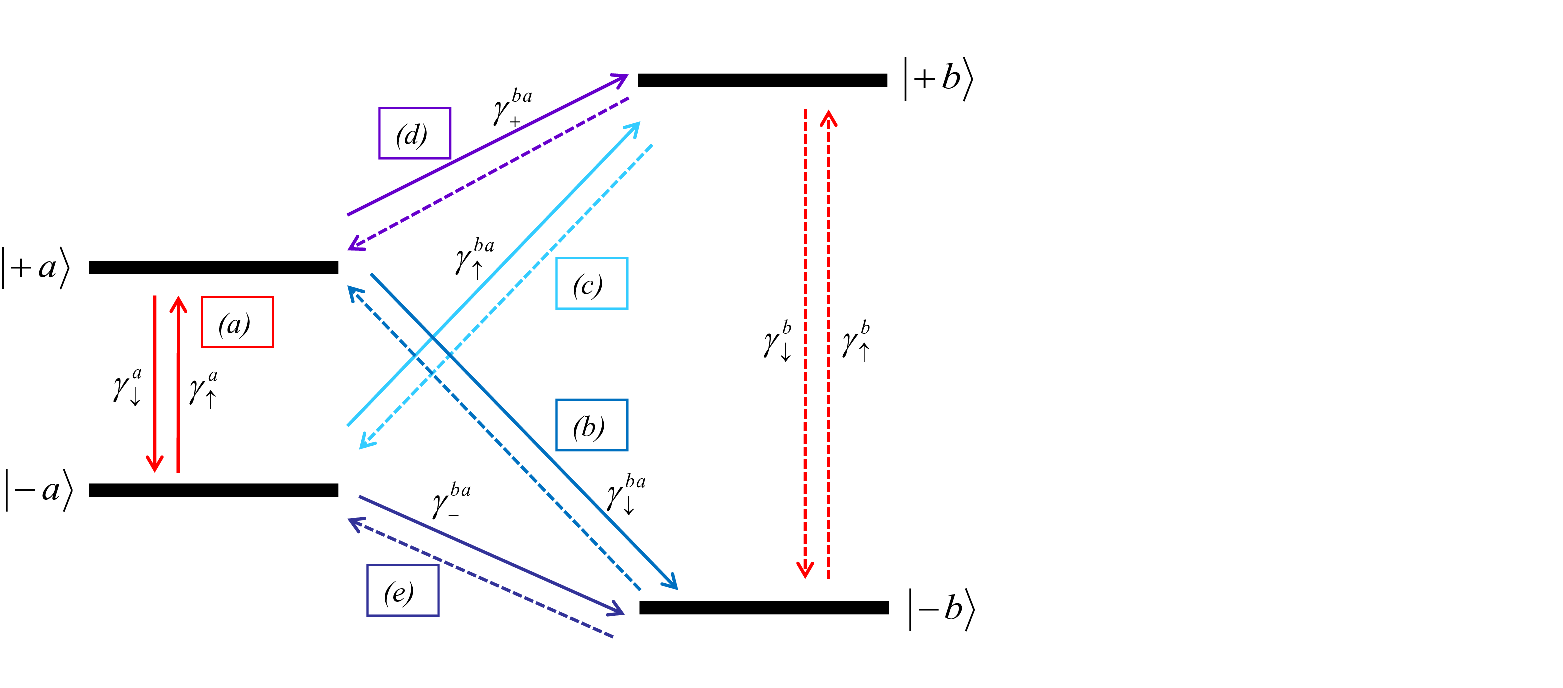}
\caption{Energy levels and coupling mechanisms associated to the evolution~(%
\protect\ref{GeneralTLS}). Each letter in the squares corresponds to each
term in this equation (see text).}
\end{figure}

\subsubsection{Minimal number of mechanisms to achieve thermality}

The detailed balance conditions~(\ref{DetBalTLS}) guaranty that the
stationary state associated to the time-evolution (\ref{GeneralTLS}) is 
\end{subequations}
\begin{equation}
\Xi _{\mathrm{th}}=w_{a}\frac{e^{-\beta \bar{H}_{a}}}{\mathrm{Tr}[e^{-\beta 
\bar{H}_{a}}]}\otimes |a\rangle \langle a|+w_{b}\frac{e^{-\beta \bar{H}_{b}}%
}{\mathrm{Tr}[e^{-\beta \bar{H}_{b}}]}\otimes |b\rangle \langle b|,
\label{ThermalTLS}
\end{equation}%
which corresponds to a hybrid quantum-classical thermal state [see Eq.~(\ref%
{ThermalPesado})]. The weights read [see Eq.~(\ref{PesosWc})]%
\begin{equation}
w_{c}=\left( \frac{e^{-\beta E_{c}}}{Z}\right) \left( \frac{\mathrm{Tr}%
[e^{-\beta \bar{H}_{c}}]}{\sum_{c^{\prime }=a,b}\frac{e^{-\beta E_{c^{\prime
}}}}{Z}\mathrm{Tr}[e^{-\beta \bar{H}_{c^{\prime }}}]}\right) ,
\label{Wc_Clasicol}
\end{equation}%
where $Z=\sum_{c=a,b}e^{-\beta E_{c}}.$ An uncorrelated solution [Eq.~(\ref%
{ThermalSeparable})] emerges when $\bar{H}_{c}\rightarrow H_{s},$ that is, $%
\bar{H}_{a}=\bar{H}_{b}=H_{s}.$

A relevant question that the present model clarify is how many mechanisms
are necessary to achieve thermalization. Taking Eq.~(\ref{GeneralTLS}) with $%
(d/dt)\rho ^{c}=0$ it is possible to check (analytically) that the thermal
state~(\ref{ThermalTLS}) can be reached without involving the five
mechanisms of Fig.~1, equivalently the corresponding contributions in Eq.~(%
\ref{GeneralTLS}). This is a general property of Lindblad equations. In
fact, thermalization is also achieved when considering a linear combination
of Lindblad dynamics, each one obeying detailed balance~\cite{alicki}.

For the studied example, taking mechanism (a) [local thermal contribution $%
\mathcal{L}_{\mathrm{th}}^{(c)}$] and any other of the remaining four
mechanisms [(b) or (c) or (d) or (e)] is enough to guarantee a stationary
thermal state. While it is not possible to establishing a general
conclusion, in this example the degeneracy on the possible mechanisms can be
associated to the dimensionality of the both subsystems and their hybrid\
quantum-classical coupling structure. If one demand the validity of detailed
balance on all possible transitions, all mechanisms sketched in Fig.~1 must
be present. On the other hand, if one impose that Eq.~(\ref{GeneralTLS})
admits, in the limit $\bar{H}_{c}\rightarrow H_{s},$ a separable solution at 
\textit{all times} (both subsystems evolves independently of each other),
added to mechanism (a), only the mechanisms (d) and (e) must be present
together.

\subsubsection{A particular example of hybrid thermal state}

As a specific case, we consider the Hamiltonian [see Eq.~(\ref{HDico})]%
\begin{equation}
\mathcal{H}=\sum_{c=a,b}E_{c}|c\rangle \langle c|+\frac{\hbar \omega _{a}}{2}%
\sigma _{z}\otimes |a\rangle \langle a|+\frac{\hbar \omega _{b}}{2}\sigma
_{x}\otimes |b\rangle \langle b|,  \label{HZX}
\end{equation}%
where $\sigma _{k}$ are the Pauli matrices, $k=x,y,z.$ The eigenstates of
each Hamiltonian are $|\pm \rangle _{a}=|\pm \rangle $ and $|\pm \rangle
_{b}=(1/\sqrt{2})(|+\rangle \pm |-\rangle ),$ where $|\pm \rangle $ are the
eigenstates of $\sigma _{z}.$

Using that $\mathrm{Tr}[e^{-\beta \bar{H}_{c}}]=2\cosh [\beta \hbar \omega
_{c}/2]$ with $c=a,b,$ the thermal states associated to the quantum
subsystem [see Eq.~(\ref{ThermalTLS})] are%
\begin{equation}
\frac{e^{-\beta \bar{H}_{a}}}{\mathrm{Tr}[e^{-\beta \bar{H}_{a}}]}=\left( 
\begin{array}{cc}
\lbrack 1+e^{+\frac{\beta \hbar \omega _{a}}{2}}]^{-1} & 0 \\ 
0 & [1+e^{-\frac{\beta \hbar \omega _{a}}{2}}]^{-1}%
\end{array}%
\right) ,  \label{ThermalZ}
\end{equation}%
while%
\begin{equation}
\frac{e^{-\beta \bar{H}_{b}}}{\mathrm{Tr}[e^{-\beta \bar{H}_{b}}]}=\frac{1}{2%
}\left( 
\begin{array}{cc}
1 & -\tanh (\frac{\beta \hbar \omega _{b}}{2}) \\ 
-\tanh (\frac{\beta \hbar \omega _{b}}{2}) & 1%
\end{array}%
\right) .
\end{equation}%
Thus, the quantum thermal state may involve coherences when looking from a
fixed basis of the two Hamiltonians. This is an interesting aspect that is
induced by underlying total Hamiltonian~(\ref{HZX}).

On the other hand, the weights of the classical subsystem can be written as
[see Eq.~(\ref{Wc_Clasicol})]%
\begin{equation}
w_{c}=\frac{e^{-\beta E_{c}}\cosh [\beta \hbar \omega _{c}/2]}{e^{-\beta
E_{a}}\cosh [\beta \hbar \omega _{a}/2]+e^{-\beta E_{b}}\cosh [\beta \hbar
\omega _{b}/2]}.
\end{equation}%
Consequently, if $E_{b}>E_{a},$ which implies that in isolation the state $b$
is less populated than state $a$ $(e^{-\beta E_{a}}>e^{-\beta E_{b}})$ the
interaction with the quantum subsystem may alter this relation, that is when 
$e^{-\beta E_{a}}\cosh [\beta \hbar \omega _{a}/2]<e^{-\beta E_{b}}\cosh
[\beta \hbar \omega _{b}/2].$ This kind of (thermal) population inversion
due to the hybrid coupling is also a central feature that the model~(\ref%
{HDico}) allows to predict. An interesting generalization of this phenomenon
appears in the next example.

\subsection{Lattice model}

Now we consider a lattice model. Hence, the state of the classical subsystem
is labeled with an index $n=0,\pm 1,\pm 2,\cdots .$ (classical coordinate).
The quantum counterpart remains as a two-level system. As a physical ground
of the model we demand that, in absence of interaction, the quantum
subsystem achieves a thermal state at each \textquotedblleft
position\textquotedblright\ of the classical subsystem. Similarly, in
absence of interaction, the classical subsystem achieve thermalization. The
interaction between both subsystems is\ introduced in such a way that the
quantum system lost its coherences (defined with respect to its own
Hamiltonian) whenever it is \textquotedblleft translated\textquotedblright\
in position.

The total hybrid Hamiltonian is taken as $\mathcal{H}=\sum_{n}E_{n}|n\rangle
\langle n|+\sum_{n}H_{n}\otimes |n\rangle \langle n|,$ where $\{E_{n}\}$
define the energy of the classical subsystem at each \textquotedblleft
site\textquotedblright\ associated to the projectors $\{|n\rangle \langle
n|\}.$ Under the required conditions, denoting with $\rho ^{n}$ the
conditional state of the quantum subsystem \textquotedblleft given
that\textquotedblright\ the classical subsystem is at site\ $n,$ we write
the time-evolution%
\begin{eqnarray}
\frac{d\rho ^{n}}{dt} &=&-i[H_{n},\rho ^{n}]+\mathcal{L}_{\mathrm{th}%
}^{(n)}[\rho ^{n}]  \label{Discrete} \\
&&+\sum_{\substack{ \tilde{n}=n\pm 1  \\ s=\pm 1}}\left( \gamma _{s}^{n%
\tilde{n}}\Pi ^{s}\rho ^{\tilde{n}}\Pi ^{s}-\frac{1}{2}\gamma _{s}^{\tilde{n}%
n}\left\{ \Pi ^{s},\rho ^{n}\right\} _{+}\right) .\ \ \ \ \ \ \   \notag
\end{eqnarray}%
The Hamiltonian of the quantum subsystem at each position\ is taken as%
\begin{equation*}
H_{n}=\frac{\hbar \omega _{n}}{2}\sigma _{z},
\end{equation*}%
where $\sigma _{z}$ is the $z$-Pauli operator. Thus, introducing the
eigenbasis $H_{n}|\pm \rangle =\varepsilon _{\pm }^{(n)}|\pm \rangle ,$ the
eigenvalues are $\varepsilon _{\pm }^{(n)}=\pm \hbar \omega _{n}/2.$ The
thermal Lindblad contributions is%
\begin{eqnarray}
\mathcal{L}_{\mathrm{th}}^{(n)}[\rho ] &=&\gamma _{\downarrow }^{n}(\sigma
\rho \sigma ^{\dagger }-\frac{1}{2}\{\sigma ^{\dagger }\sigma ,\rho \}_{+}) 
\notag \\
&&+\gamma _{\uparrow }^{n}(\sigma ^{\dagger }\rho \sigma -\frac{1}{2}%
\{\sigma \sigma ^{\dagger },\rho \}_{+}),  \label{LthEne}
\end{eqnarray}%
where the raising and lowering operators are $\sigma ^{\dagger }=|+\rangle
\langle -|$ and $\sigma =|-\rangle \langle +|$ respectively. The rates
fulfill%
\begin{equation}
\frac{\gamma _{\uparrow }^{n}}{\gamma _{\downarrow }^{n}}=e^{-\beta \lbrack
\varepsilon _{+}^{(n)}-\varepsilon _{-}^{(n)}]}.  \label{BalanceDiagLattice}
\end{equation}%
On the other hand, the two non-diagonal mechanisms, which couple first
neighbor states ($\rho ^{n}\leftrightarrow \rho ^{n\pm 1}$), are defined by
the projectors $\Pi _{\pm }=|\pm \rangle \langle \pm |.$ The corresponding
rates fulfill%
\begin{equation}
\frac{\gamma _{\pm }^{n\tilde{n}}}{\gamma _{\pm }^{\tilde{n}n}}=e^{-\beta
\lbrack E_{n}+\varepsilon _{\pm }^{(n)}-(E_{\tilde{n}}+\varepsilon _{\pm }^{(%
\tilde{n})})]}.  \label{DetBalLattice}
\end{equation}%
%
%
%
%
%
%
%
%
%
%
%
%
%
%
%
%
%
%
%
%
%
%
%
%
%
%
%
%
%
%
%
%
%
%
%
%
%
%
%
%
%
%
%
%
%
%
%
%
%
%
%
%
%
%
%
%
%
\begin{figure}[tbp]
\includegraphics[bb=65 15 675 350,
angle=0,width=8.5cm]{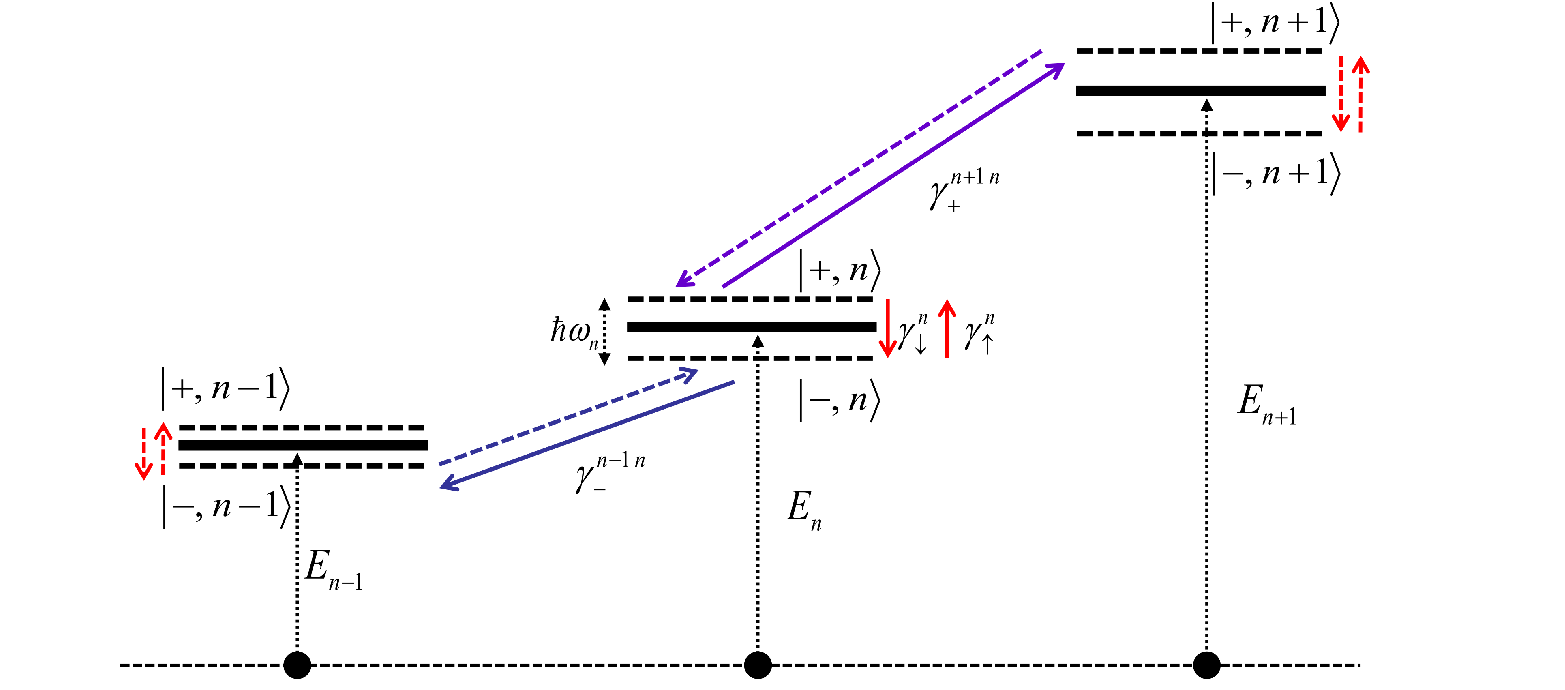}
\caption{Coupling mechanisms associated to the evolution~(\protect\ref%
{Discrete}). The energy levels approximately represent the dependences
defined by Eq.~(\protect\ref{Energias}).}
\end{figure}

In Fig.~2 we sketch the energy levels and coupling mechanisms corresponding
to the evolution~(\ref{Discrete}). The transitions $|+,n\rangle \overset{%
\gamma _{\downarrow }^{n}}{\rightarrow }|-,n\rangle $ and $|-,n\rangle 
\overset{\gamma _{\uparrow }^{n}}{\rightarrow }|+,n\rangle $ correspond to
the isolated action of $\mathcal{L}_{\mathrm{th}}^{(n)}.$ Taking into
account the detailed balance condition~(\ref{BalanceDiagLattice}), as
required, these transitions tend to thermalize the quantum subsystem at each 
$n$-position. The transitions $|+,n\rangle \overset{\gamma _{+}^{n\pm 1,n}}{%
\rightarrow }|+,n\pm 1\rangle $ and $|-,n\rangle \overset{\gamma _{-}^{n\pm
1,n}}{\rightarrow }|-,n\pm 1\rangle $ [equivalent to (d) and (e) in Fig.~1]
change the state of the classical subsystem, $n\rightarrow n\pm 1,$ and
represent two dephasing mechanisms for the quantum subsystem. The extra
detailed balance condition~(\ref{DetBalLattice}) guarantee that the
arrangement achieves a hybrid thermal state in the stationary regime.

In Appendix~\ref{MatrixElements}\ we provide the time-evolution of the
matrix elements of $\rho ^{n}.$ Consistently with Eq.~(\ref{ThermalPesado}),
it is simple to cheek that the stationary state associated to Eq.~(\ref%
{Discrete}) is%
\begin{equation}
\Xi _{\mathrm{th}}=\sum_{n=-\infty }^{+\infty }w_{n}\frac{e^{-\beta H_{n}}}{%
\mathrm{Tr}[e^{-\beta H_{n}}]}\otimes |n\rangle \langle n|,
\label{ThermalLattice}
\end{equation}%
where $e^{-\beta H_{n}}/\mathrm{Tr}[e^{-\beta H_{n}}]$ can be read from Eq.~(%
\ref{ThermalZ}) under the replacement $\omega _{a}\rightarrow \omega _{n}.$
The weights are%
\begin{equation}
w_{n}=\frac{e^{-\beta E_{n}}}{Z}\left( \frac{\cosh [\beta \hbar \omega
_{n}/2].}{\sum_{\tilde{n}=-\infty }^{+\infty }\frac{e^{-\beta E_{\tilde{n}}}%
}{Z}\cosh [\beta \hbar \omega _{\tilde{n}}/2].}\right) ,\ \ \ 
\label{WnLattice}
\end{equation}%
with $Z=\sum_{n=-\infty }^{+\infty }e^{-\beta E_{n}}$ and we used that $%
\mathrm{Tr}[e^{-\beta H_{n}}]=2\cosh [\beta \hbar \omega _{n}/2].$

In Appendix~\ref{AlternativeLattice} we present a lattice model based on the
coupling mechanisms~(b) and~(c) of Fig.~1. They lead to alternative
dephasing mechanisms which in turn also induce, in the stationary regime, to
the same hybrid thermal state, Eq.~(\ref{ThermalLattice}).

\subsubsection{Energy dependences}

The model is completely specified after providing the dependence on each
site of the energy of both subsystems. We notice that in different physical
platforms by applying an external (extended) field it is simple to modify
the characteristic energy of a quantum system in a linear way (in a given
coordinate). Furthermore, a harmonic classical system corresponds to a
quadratic dependence of energy with position.

With the previous motivation, we assume the dependences%
\begin{equation}
\hbar \omega _{n}=\hbar (\omega _{0}+\delta \omega |n|),\ \ \ \ \ \ \ \
E_{n}=E_{0}+\delta E\text{\thinspace }n^{2}.  \label{Energias}
\end{equation}%
Here, $\delta \omega $ measures the difference of \textquotedblleft quantum
energy\textquotedblright\ between neighbors sites while $\delta E$ measures
the difference of \textquotedblleft classical energy.\textquotedblright\
From Fig.~2, taking into account the general result~(\ref{PesosWc}), we
realize that under the condition $\hbar \delta \omega \gg \delta E$ the
thermal distribution of the classical system must be strongly modified. In
fact, the transitions associated to the rates $\gamma _{+}^{n\pm 1,n}$ and $%
\gamma _{-}^{n\pm 1,n}$ gives two alternatives way for achieving
equilibrium. The difference of energy between neighbor states $(|\pm ,n\pm
1\rangle $ and $|\pm ,n\rangle )$ is $[(E_{n\pm 1}\pm \hbar \omega
_{n}/2)-(E_{n}\pm \hbar \omega _{n}/2)]\approx \delta E[n\pm \mathrm{sgn}%
(n)\hbar \delta \omega /(4\delta E$\thinspace $)],$ where $\mathrm{sgn}(x)$\
is the sign function. This difference of energy can be read as a linear
force (in the variable $n)$ that in contrast to the isolated case is shifted
by $\pm \hbar \delta \omega /(4\delta E).$\thinspace Thus, under the
condition $\hbar \delta \omega \gg \delta E$ we expect a bimodal behavior
for the distribution of the classical subsystem.

\subsubsection{Continuous (smooth) limit and bimodal behavior}

In order to check the previous qualitative argument, instead of
characterizing the discrete case, in addition we assume that%
\begin{equation}
\beta \delta E\ll 1.  \label{ConditionSuave}
\end{equation}%
Under this condition a smooth continuous limit approximation applies. Hence,
we introduce a continuous coordinate $x=\delta xn,$ where $\delta x$
measures the \textquotedblleft distance\textquotedblright\ between $n$ and $%
n\pm 1.$ In the limit $\beta \delta E\ll 1,$ the isolated thermal
distribution of the classical subsystem can\ be very well approximated as%
\begin{equation}
\frac{e^{-\beta E_{n}}}{Z}\simeq G_{\mathrm{th}}(x)\delta x\equiv \left( 
\sqrt{\frac{\beta \delta E}{\pi \delta x^{2}}}\exp \left[ -\frac{\beta
\delta E}{\delta x^{2}}x^{2}\right] \right) \delta x.  \label{Gauss}
\end{equation}%
This equivalence can be understood by checking that, to first order in $%
\delta x,$ the approximation $e^{-\beta E_{n\pm 1}}/Z\simeq G_{\mathrm{th}%
}(x\pm \delta x)\delta x$ is valid under the condition~(\ref{ConditionSuave}%
).

Eq.~(\ref{Gauss}) implies that, as expected, when isolated the classical
subsystem is characterized by a Gaussian thermal distribution,
characteristic of a harmonic system. Similarly, the thermal bipartite state~(%
\ref{ThermalLattice}) is written as%
\begin{equation}
\Xi _{\mathrm{th}}=\int_{-\infty }^{+\infty }dxw(x)\frac{e^{-\beta H_{x}}}{%
\mathrm{Tr}[e^{-\beta H_{x}}]}\otimes |x\rangle \langle x|.
\label{ThermalContinuo}
\end{equation}%
Here, the weight $w(x)$ becomes a probability density that reads%
\begin{equation}
w(x)=\frac{1}{\mathcal{Z}_{\mathrm{th}}}G_{\mathrm{th}}(x)\cosh \left[ \frac{%
1}{2}\left( \beta \hbar \omega _{0}+\frac{\beta \hbar \delta \omega }{\delta
x}|x|\right) \right] ,  \label{PesoX}
\end{equation}%
where as before [Eq.~(\ref{WnLattice})], the second factor corresponds to
the quantum partition function at each site. The normalization constant, by
integration, is $\mathcal{Z}_{\mathrm{th}}=\exp [\frac{\beta (\hbar \delta
\omega )^{2}}{16\delta E}][\cosh (\frac{\beta \hbar \omega _{0}}{2})+\mathrm{%
erf}(\frac{\beta \hbar \delta \omega }{4\sqrt{\beta \delta E}})\sinh (\frac{%
\beta \hbar \omega _{0}}{2})],$ where $\mathrm{erf}(x)$ is the error
function.

In Fig.~3 we plot the density $w(x)$ as a function of the coordinate $x$ and
for different values of $\hbar \delta \omega /\delta E$ and $\beta \hbar
\omega _{0}=0.$ When $\hbar \delta \omega /\delta E\lesssim 1$ the density $%
w(x),$ in the scale of the plot, is indistinguishable from the
non-interacting case, that is, from the Gaussian density $G_{\mathrm{th}}(x)$
[Fig.~2(a)]. When $\hbar \delta \omega /\delta E\gtrsim 1$ departures with
respect to the Gaussian case are developed [Fig. 2(b)]. When $\hbar \delta
\omega /\delta E\gg 1,$ $w(x),$ as expected, develops a bimodal behavior
[Fig. 2(c) and (d)]. 
\begin{figure}[tbp]
\includegraphics[bb=45 590 740 1153,
angle=0,width=8.5cm]{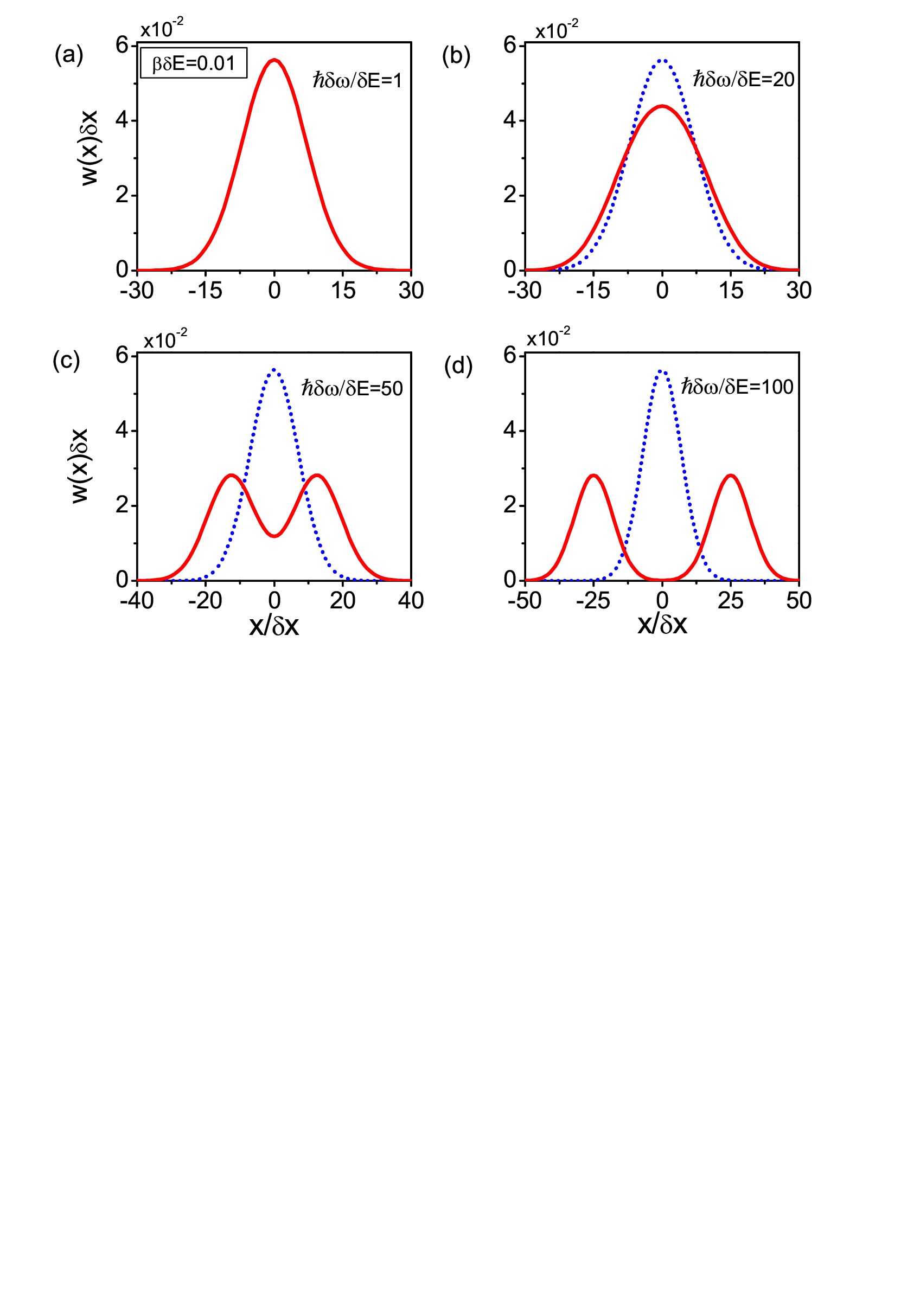}
\caption{Probability density $w(x)$ [Eq.~(\protect\ref{PesoX})]\ of the
classical subsystem as a function of the coordinate $x.$ The characteristic
parameters are $\protect\beta \protect\delta E=0.01$ and $\protect\beta %
\hbar \protect\omega _{0}=0.$ The value of $\hbar \protect\delta \protect%
\omega /\protect\delta E$ is indicated in each plot. In all cases, the
dotted lines correspond to the uncorrelated case where $w(x)=G_{\mathrm{th}%
}(x)$ [Eq.~(\protect\ref{Gauss})].}
\end{figure}
The bimodal behavior is induced by the influence of the quantum partition
function in Eq.~(\ref{PesoX}). This factor reflects how the thermodynamics
of the classical subsystem is affected by the quantum one and has its origin
in the general result expressed by Eq.~(\ref{PesosWc}) and the energies
dependences~(\ref{Energias}).

Under the identifications $w(x)\delta x\rightarrow w_{n}$ and $x/\delta
x\rightarrow n,$ the curves of Fig.~3 are indistinguishable (in the scale of
the plots) with the discrete solutions set by Eq.~(\ref{WnLattice}). This
equivalence is granted by condition~(\ref{ConditionSuave}) and demonstrates
that the main features shown in Fig.~2 do not depend on the applicability of
the smooth approximation. In fact, as said before, the bimodal behavior is a
consequence of Eq.~(\ref{PesosWc}) and the energy dependences introduced
through Eq.~(\ref{Energias}). Nevertheless, the physical origin of this
behavior can be characterized in an alternative complementary way by
studying the time-evolution [Eq.~(\ref{Discrete})] in the same continuous
limit.

\subsubsection{Quantum Fokker-Planck-like equation}

Writing the bipartite state as $\Xi =\int_{-\infty }^{+\infty }dx\varrho
_{x}\otimes |x\rangle \langle x|,$ using a standard diffusive approximation~%
\cite{kampen,budini}, the time-evolution of the conditional (density) state $%
\varrho _{x}$ reads%
\begin{eqnarray}
\frac{d\varrho _{x}}{dt} &\simeq &-i[H_{x},\varrho _{x}]+\mathcal{L}_{%
\mathrm{th}}^{(x)}[\varrho _{x}]  \notag \\
&&+\sum_{s=\pm 1}\gamma \left( \Pi ^{s}\varrho _{x}\Pi ^{s}-\frac{1}{2}%
\left\{ \Pi ^{s},\varrho _{x}\right\} _{+}\right)  \label{LatticeContinuas}
\\
&&+\sum_{s=\pm 1}\gamma \mathcal{L}_{\mathrm{FP}}^{(s)}[\Pi ^{s}\varrho
_{x}\Pi ^{s}],  \notag
\end{eqnarray}%
where the Fokker-Planck operator is defined as%
\begin{equation}
\mathcal{L}_{\mathrm{FP}}^{(\pm )}[\bullet ]=\left\{ -\beta \frac{\partial }{%
\partial x}[f_{\pm }(x)(\bullet )]+\frac{\partial ^{2}(\bullet )}{\partial
^{2}x}\right\} \frac{\delta x^{2}}{2}.  \label{Lfp}
\end{equation}%
In each case, the force reads%
\begin{equation}
f_{\pm }(x)=-2\left[ \frac{\delta E}{\delta x^{2}}x\pm \mathrm{sgn}(x)\frac{%
\hbar \delta \omega }{4\delta x}\right] .
\end{equation}

In Appendix~\ref{MatrixElements} the time-evolution of the matrix elements
of $\varrho _{x}$ are explicitly written. The structure of Eq.~(\ref%
{LatticeContinuas}) can be easily understood from Eq.~(\ref{Discrete}). In
both expressions the first lines define the local Hamiltonian and local
thermalization at each site. The second and third lines in Eq.~(\ref%
{LatticeContinuas}) corresponds to a diffusive approximation of the hybrid
coupling mechanism defined by the second line in Eq.~(\ref{Discrete}). Their
derivation is sketched in Appendix~\ref{QFPApprox}. Notice that the second
line in Eq.~(\ref{LatticeContinuas}) introduces local dephasing mechanisms,
while the third line is set by the Fokker-Planck contributions $\mathcal{L}_{%
\mathrm{FP}}^{(\pm )}[\bullet ].$ The rate $\gamma $ represents the degree
of freedom lefts by the detailed balance condition [Eq.~(\ref{DetBalLattice}%
)] fulfilled by the rates. Furthermore, as usual in a continuous diffusive
approximation~\cite{barkai}, consistently with the conditions~(\ref%
{DetBalLattice}) and~(\ref{ConditionSuave}), we approximated $\gamma _{\pm
}^{n+1,n}+\gamma _{\pm }^{n-1,n}\simeq \gamma $ and $\gamma _{\pm
}^{n+1,n}-\gamma _{\pm }^{n-1,n}\simeq \gamma \beta f_{\pm }(x)\delta x/2.$

The underlying physical origin of the forces $f_{\pm }(x)$ can be related
with the symmetrical dependence of the frequency of the quantum subsystem
with position, Eq.~(\ref{Energias}). The change in sign $(\pm )$ takes into
account that the projectors $\Pi ^{+}$ and $\Pi ^{-}$ in Eq.~(\ref{Discrete}%
) only couple upper with upper states and lower with lower states (see
Fig.~2), which has associated respectively the changes of energy $\pm \hbar
\delta \omega /2.$

After some standard algebra, it is possible to check that the state~(\ref%
{ThermalContinuo}) defines the stationary solution of the Fokker-Planck-like
evolution~(\ref{LatticeContinuas}). This time-evolution allows us to
understand the origin\ of the bimodal behavior from a dynamical perspective.
In fact, the forces defined by $f_{\pm }(x)$ correspond to the potential
energies%
\begin{equation}
V_{\pm }(x)=\delta E\left( \frac{x}{\delta x}\pm \mathrm{sgn}(x)\frac{\hbar
\delta \omega }{4\delta E}\right) ^{2},  \label{ShiftPot}
\end{equation}%
obtained from $f_{\pm }(x)=-(\partial /\partial x)V_{\pm }(x).$ Hence, the
potential energy corresponds to an harmonic potential whose minimal value is
shifted by the quantum-classical interaction. Consistently with the previous
qualitative arguments, this shifting is proportional to the product of $%
(\hbar \delta \omega /4\delta E),$ property that supports the emergence, in
the stationary regime, of the bimodal behavior shown in Fig.~3 when $(\hbar
\delta \omega /\delta E)\gg 1.$

\section{Summary and conclusions}

A full characterization of thermal states when considering quantum and
classical interacting subsystems has been presented. The developed approach
relies on embedding the hybrid arrangement in a full quantum bipartite
description [Eq.~(\ref{BipartiteState})].

Considering a canonical ensemble we characterized how the thermal states of
each subsystem is affected by the interaction with the other one [Eq.~(\ref%
{ThermalPesado})]. The (partial) state of the quantum system becomes a
statistical superposition of the thermal states associated to each
Hamiltonian related to each state of the classical subsystem [Eq.~(\ref%
{RhoThS})]. In turn, the thermal state of the classical subsystem is
modified by a multiplicative term that depends on the Helmholtz free energy
associated to each quantum conditional thermal state [Eq.~(\ref{PesosWc})].

As a main result we established which kind of time evolutions leads in the
stationary (long time) regime to a hybrid thermal state. The approach relies
on standard Lindblad equations that fulfill a detailed balance condition.
Under these assumptions, the evolution of the conditional quantum states is
set by a hybrid Lindblad equation [Eq.~(\ref{ThermalHybrid})] which in turn
can be read as a collisional dynamics where each event introduces a change
of eigenbasis corresponding to the underlying Hamiltonians [Eq.~(\ref{GenUU}%
)].

The full approach was exemplified through a set of paradigmatic examples.
Considering that both subsystems are two-level systems [Eq.~(\ref{GeneralTLS}%
)], we established that different coupling mechanisms [Fig.~1] could lead to
the same stationary hybrid thermal state. In addition, considering a lattice
model [Eq.~(\ref{Discrete})] we characterized how a Gaussian thermal state
associated to a classical harmonic system could become a bimodal
distribution when increasing the interaction with a quantum two-level
subsystem [Fig.~3]. This property was also understood from a shifting of the
quadratic potential of the classical subsystem induced by the
(time-irreversible) coupling with the quantum subsystem [Eq.~(\ref{ShiftPot}%
)].

Our approach, under the assumptions of local-in-time evolutions added to a
detailed balance condition, constraint the possible time-evolutions that
lead in the stationary limit to a quantum-classical thermal state. Added to
its intrinsic theoretical interest, the developed results could trigger
extra research on related characteristics of hybrid systems and dynamics.
For example, given that the partial dynamics of each subsystem (the
classical or the quantum one) depends on extra degrees of freedom, the
underlying model could be the starting point to studying non-Markovian
effects~\cite{entropy} under a hybrid thermalization condition. Furthermore,
as the hybrid dynamics can be derived from a full microscopic unitary
dynamics~\cite{rate} the quantum subsystem dynamics can also be read as an
approximation in a strong coupling regime with the (full) environment.
Hence, recent progress on the strong coupling thermodynamics of open quantum
systems~\cite{Qthermo,asia,angel} could be analyzed in the context of the
developed model. Finally, it is interesting to note that hybrid
quantum-classical dynamics also emerges in the study of nonadiabatic
molecular dynamics~\cite{tully,kapral,crespo}. The developed results could
in principle be used as an alternative starting frame to study
thermalization in these kinds of systems.

\section*{Acknowledgments}

I thanks to Prof. Jos\'{e} Luiz Alonso, Universidad de Zaragoza, Spain, for
drawing my attention about the problem of thermalization in hybrid
quantum-classical dynamics. Support from Consejo Nacional de Investigaciones
Cient\'{\i}ficas y T\'{e}cnicas (CONICET), Argentina, is also acknowledged.

\appendix

\section{Derivation of the thermal hybrid evolution \label{Derivation}}

Given the hybrid bipartite Hamiltonian $\mathcal{H}=\sum_{c}H_{c}\otimes
|c\rangle \langle c|,$ its eigensystem is denoted as $\mathcal{H}|j,c\rangle
=\varepsilon _{j}^{(c)}|j,c\rangle $ where $\varepsilon _{j}^{(c)}$ is the $%
j $-eigenvalue of the Hamiltonian $H_{c}.$ Based on the results of Sec.~(\ref%
{ThermalSec}), each pair of eigenvalues of the Hamiltonian here becomes $%
\{i,j\}\rightarrow \{ic,\tilde{j}\tilde{c}\}.$ We consider in a separate way
the cases \textit{(i)} $\{ic,jc\}$ with $i\neq j,$ and \textit{(ii)} $\{ic,%
\tilde{j}\tilde{c}\}$ with $c\neq \tilde{c}.$

In the first case, $\{ic,jc\}$ with $i\neq j,$ from Eq.~(\ref{Lij}), taking
into account the preservation of the state~(\ref{BipartiteState}), we write
the Lindblad contributions%
\begin{eqnarray}
\mathcal{L}_{\{i,j\}}^{(c)}[\rho ^{c}] &=&\gamma _{ij}^{(c)}\left( |i\rangle
\langle j|\rho ^{c}|j\rangle \langle i|-\frac{1}{2}\left\{ |j\rangle \langle
j|,\rho ^{c}\right\} _{+}\right)  \notag \\
&&+\gamma _{ji}^{(c)}\left( |j\rangle \langle i|\rho ^{c}|i\rangle \langle
j|-\frac{1}{2}\left\{ |i\rangle \langle i|,\rho ^{c}\right\} _{+}\right) ,\
\ \ \ \ \   \label{Diagonal_IJ}
\end{eqnarray}%
where $\rho ^{c}=\langle c|\Xi |c\rangle $ [Eq.~(\ref{BipartiteState})]. The
sum $\mathcal{L}_{\mathrm{th}}^{(c)}\equiv \sum_{\{i,j\}}\mathcal{L}%
_{\{i,j\}}^{(c)}$ is a standard Lindblad equation whose isolated action
leads to the thermal state associated to the Hamiltonian $H_{c}.$ The rates
appearing in each $\mathcal{L}_{\mathrm{th}}^{(c)}$ must fulfill condition~(%
\ref{ThermalRates}), which leads to Eq.~(\ref{DiagonalThermalRates}).

In the second case, $\{ic,\tilde{j}\tilde{c}\}$ with $c\neq \tilde{c},$ from
Eq.~(\ref{Lij}) it follows%
\begin{eqnarray*}
\mathcal{L}_{\{ic,\tilde{j}\tilde{c}\}}[\rho ]\! &=&\!\gamma _{i\tilde{j}%
}^{(c\tilde{c})}\left( |ic\rangle \langle \tilde{j}\tilde{c}|\rho |\tilde{j}%
\tilde{c}\rangle \langle ic|-\frac{1}{2}\left\{ |\tilde{j}\tilde{c}\rangle
\langle \tilde{j}\tilde{c}|,\rho \right\} _{+}\right) \\
&&\!\!+\gamma _{\tilde{j}i}^{(\tilde{c}c)}\left( |\tilde{j}\tilde{c}\rangle
\langle ic|\rho |ic\rangle \langle \tilde{j}\tilde{c}|-\frac{1}{2}\left\{
|ic\rangle \langle ic|,\rho \right\} _{+}\right) .
\end{eqnarray*}%
Taking into account the state~(\ref{BipartiteState}), we write%
\begin{eqnarray*}
\mathcal{L}_{\{ic,\tilde{j}\tilde{c}\}}[\rho ]\! &=&\!\gamma _{i\tilde{j}%
}^{(c\tilde{c})}\left( |ic\rangle \langle \tilde{j}|\rho ^{\tilde{c}}|\tilde{%
j}\rangle \langle ic|-\frac{1}{2}\left\{ |\tilde{j}\tilde{c}\rangle \langle 
\tilde{j}\tilde{c}|,\rho ^{\tilde{c}}\right\} _{+}\right) \\
&&\!\!+\gamma _{\tilde{j}i}^{(\tilde{c}c)}\left( |\tilde{j}\tilde{c}\rangle
\langle i|\rho ^{c}|i\rangle \langle \tilde{j}\tilde{c}|-\frac{1}{2}\left\{
|ic\rangle \langle ic|,\rho ^{c}\right\} _{+}\right) .
\end{eqnarray*}%
Thus, we get the contribution%
\begin{equation}
\mathcal{L}_{\{ic,\tilde{j}\tilde{c}\}}[\rho ^{c}]\!=\!-\frac{1}{2}\gamma _{%
\tilde{j}i}^{(\tilde{c}c)}\left\{ |i\rangle \langle i|,\rho ^{c}\right\}
_{+}+\gamma _{i\tilde{j}}^{(c\tilde{c})}|i\rangle \langle \tilde{j}|\rho ^{%
\tilde{c}}|\tilde{j}\rangle \langle i|,  \label{Uno}
\end{equation}%
and symmetrically%
\begin{equation}
\mathcal{L}_{\{ic,\tilde{j}\tilde{c}\}}[\rho ^{\tilde{c}}]\!=\!-\frac{1}{2}%
\gamma _{i\tilde{j}}^{(c\tilde{c})}\left\{ |\tilde{j}\rangle \langle \tilde{j%
}|,\rho ^{\tilde{c}}\right\} _{+}+\gamma _{\tilde{j}i}^{(\tilde{c}c)}\left( |%
\tilde{j}\rangle \langle i|\rho ^{c}|i\rangle \langle \tilde{j}|\right) .
\label{Dos}
\end{equation}%
These two last equations represent the transitions $|ic\rangle \rightarrow |%
\tilde{j}\tilde{c}\rangle $ and $|ic\rangle \leftarrow |\tilde{j}\tilde{c}%
\rangle $ respectively. Similarly to the diagonal case, the rates must
fulfill condition~(\ref{ThermalRates}).

The contributions defined by the diagonal contribution Eq.~(\ref{Diagonal_IJ}%
) and the non-diagonal contributions Eqs.~(\ref{Uno}) and~(\ref{Dos}) lead to%
\begin{eqnarray}
\frac{d\rho ^{c}}{dt} &=&\mathcal{L}_{\mathrm{th}}^{(c)}[\rho ^{c}]-\frac{1}{%
2}\sum_{\substack{ \tilde{c}  \\ \tilde{c}\neq c}}\sum_{\{\tilde{j}%
,i\}}\gamma _{\tilde{j}i}^{(\tilde{c}c)}\left\{ |i\rangle \langle i|,\rho
^{c}\right\} _{+}  \notag \\
&&+\sum_{\substack{ \tilde{c}  \\ \tilde{c}\neq c}}\sum_{\{\tilde{j}%
,i\}}\gamma _{i\tilde{j}}^{(c\tilde{c})}|i\rangle \langle \tilde{j}|\rho ^{%
\tilde{c}}|\tilde{j}\rangle \langle i|.  \label{FinalThermal}
\end{eqnarray}%
The general evolution defined by Eq.~(\ref{ThermalHybrid}) follows from Eq.~(%
\ref{FinalThermal}) after expressing all coupling contributions in terms of
the transitions operators~(\ref{Aes}).

\section{Bipartite embedding \label{embedding}}

Let a bipartite density matrix evolves as%
\begin{equation}
\frac{d\Xi }{dt}=\sum_{\mu }\eta _{\mu }(V_{\mu }\Xi V_{\mu }^{\dag }-\frac{1%
}{2}\{V_{\mu }^{\dag }V_{\mu },\Xi \}_{+}).  \label{Lindblad}
\end{equation}%
Take only two operators\ $(\mu =a,b)$ defined as%
\begin{equation}
V_{a}=A\otimes |\tilde{c}\rangle \langle c|,\ \ \ \ \ \ \ V_{b}=B^{\dagger
}\otimes |c\rangle \langle \tilde{c}|.
\end{equation}%
Assuming the structure $\Xi =\sum_{c}\rho ^{c}\otimes |c\rangle \langle c|,$
the evolution of the auxiliary states reads%
\begin{eqnarray}
\frac{d\rho ^{c}}{dt} &=&-\frac{1}{2}\eta _{a}\left\{ A^{\dagger }A,\rho
^{c}\right\} _{+}+\eta _{b}B^{\dagger }\rho ^{\tilde{c}}B, \\
\frac{d\rho ^{\tilde{c}}}{dt} &=&-\frac{1}{2}\eta _{b}\left\{ BB^{\dagger
},\rho ^{c}\right\} _{+}+\eta _{a}A\rho ^{c}A^{\dagger }.
\end{eqnarray}%
When the evolution is diagonal%
\begin{equation}
\frac{d\rho ^{c}}{dt}=\eta _{a}\left( -\frac{1}{2}\left\{ A^{\dagger }A,\rho
^{c}\right\} _{+}+A\rho ^{c}A^{\dagger }\right) ,
\end{equation}%
the operator in Eq.~(\ref{Lindblad}) must be $V_{\mu }\rightarrow A\otimes
|c\rangle \langle c|.$

The previous expressions allow us to build up an explicit bipartite Lindblad
equation associated to the hybrid evolution~(\ref{ThermalHybrid}). The
bipartite density matrix $\Xi $ evolves as in Eq.~(\ref{Lindblad}) but where
the operators corresponding to the non-diagonal contributions are%
\begin{equation}
V_{\mu }\rightarrow A_{\tilde{j}i}\otimes |\tilde{c}\rangle \langle c|,\ \ \
and\ \ \ \ V_{\mu }\rightarrow A_{i\tilde{j}}\otimes |c\rangle \langle 
\tilde{c}|,
\end{equation}%
while the diagonal contributions correspond to the operators%
\begin{equation}
V_{\mu }\rightarrow A_{ji}\otimes |c\rangle \langle c|.
\end{equation}

\section{Evolution of the density matrix elements \label{MatrixElements}}

Defining the matrix elements $p_{n}^{(\pm )}\equiv \langle \pm |\rho
_{n}|\pm \rangle ,$ from Eq.~(\ref{Discrete}) it follows the evolutions 
\begin{subequations}
\label{pnEvolution}
\begin{eqnarray}
\frac{dp_{n}^{(+)}}{dt} &=&-\gamma _{\downarrow }^{n}p_{n}^{(+)}+\gamma
_{\uparrow }^{n}p_{n}^{(-)}+\mathrm{L}_{\mathrm{ME}}^{(+)}[\{p_{n}^{(+)}\}],
\\
\frac{dp_{n}^{(-)}}{dt} &=&-\gamma _{\uparrow }^{n}p_{n}^{(-)}+\gamma
_{\downarrow }^{n}p_{n}^{(+)}+\mathrm{L}_{\mathrm{ME}}^{(-)}[\{p_{n}^{(-)}%
\}],
\end{eqnarray}%
where the master-equation operator $\mathrm{L}_{\mathrm{ME}%
}^{(s)}[\{p_{n}\}] $ only couples first neighbors. It reads $(s=\pm 1)$%
\end{subequations}
\begin{equation}
\mathrm{L}_{\mathrm{ME}}^{(s)}[\{p_{n}\}]=\sum_{\tilde{n}=n\pm 1}\gamma
_{s}^{n\tilde{n}}p_{\tilde{n}}-\sum_{\tilde{n}=n\pm 1}\gamma _{s}^{\tilde{n}%
n}p_{n}.
\end{equation}%
For the coherences $c_{x}^{(\pm )}\equiv \langle \pm |\rho _{n}|\mp \rangle $
we get%
\begin{equation}
\frac{dc_{n}^{(\pm )}}{dt}\simeq \mp i\omega _{n}c_{n}^{(\pm )}-\frac{1}{2}%
\left( \gamma _{\downarrow }^{n}+\gamma _{\uparrow }^{n}+\sum_{\tilde{n}%
=n\pm 1}\gamma _{s}^{\tilde{n}n}\right) c_{n}^{(\pm )},  \label{cnEvolution}
\end{equation}%
whose solution only involves an exponential decay with oscillations of
frequency $\omega _{n}.$

In the continuous limit, these time-evolutions are approximated by the
matrix elements of Eq.~(\ref{LatticeContinuas}). Defining $P_{x}^{(\pm
)}\equiv \langle \pm |\varrho _{x}|\pm \rangle ,$ it follows 
\begin{subequations}
\begin{eqnarray}
\frac{dP_{x}^{(+)}}{dt} &\simeq &-\gamma _{\downarrow
}^{(x)}P_{x}^{(+)}+\gamma _{\uparrow }^{(x)}P_{x}^{(-)}+\gamma \mathcal{L}_{%
\mathrm{FP}}^{(+)}[P_{x}^{(+)}],\ \ \ \ \ \  \\
\frac{dP_{x}^{(-)}}{dt} &\simeq &-\gamma _{\uparrow
}^{(x)}P_{x}^{(-)}+\gamma _{\downarrow }^{(x)}P_{x}^{(+)}+\gamma \mathcal{L}%
_{\mathrm{FP}}^{(-)}[P_{x}^{(-)}],\ \ \ \ \ \ 
\end{eqnarray}%
while for the coherences $C_{x}^{(\pm )}\equiv \langle \pm |\varrho _{x}|\mp
\rangle $ we get 
\end{subequations}
\begin{equation}
\frac{dC_{x}^{(\pm )}}{dt}\simeq \mp i\omega _{x}C_{x}^{(\pm )}-\frac{1}{2}%
\left( \gamma _{\downarrow }^{x}+\gamma _{\uparrow }^{x}+\gamma \right)
C_{x}^{(\pm )}.
\end{equation}%
The thermal state~(\ref{ThermalContinuo}) is the stationary solution of
these equations. Nevertheless, it is not possible to guaranty a physical
behavior at all times, that is, the (density) matrix $\varrho _{x}$ may
lacks\ its positive definite character. In fact, Eq.~(\ref{LatticeContinuas}%
) does not has the structure of a quantum Fokker-Planck equation~\cite%
{oppen1,oppen2,lastDiosi,soda} were the positivity of $\varrho _{x}$ is
granted at all times. Nevertheless, as demonstrated in Ref.~\cite{budini}
these unphysical effects emerges in a short time regime. Notice that in
contrast, this problem does not emerge from the evolutions~(\ref{pnEvolution}%
) and~(\ref{cnEvolution}). In fact, the time-evolution~(\ref{Discrete}),
given that it can be alternatively be written as a bipartite Lindblad
equation [see Appendix~\ref{embedding}] physical solutions at any time
regime are granted.

\section{Alternative lattice model \label{AlternativeLattice}}

Here we consider the alternative dynamics%
\begin{eqnarray}
\frac{d\rho ^{n}}{dt} &=&-i[H_{n},\rho ^{n}]+\mathcal{L}_{\mathrm{th}%
}^{(n)}[\rho ^{n}] \\
&&+\sum_{\tilde{n}=n\pm 1}\left( \gamma _{\uparrow }^{n\tilde{n}}\sigma
^{\dagger }\rho ^{\tilde{n}}\sigma -\frac{1}{2}\gamma _{\downarrow }^{\tilde{%
n}n}\left\{ \sigma ^{\dagger }\sigma ,\rho ^{n}\right\} _{+}\right) \ \ \ \ 
\notag \\
&&+\sum_{\tilde{n}=n\pm 1}\left( \gamma _{\downarrow }^{n\tilde{n}}\sigma
\rho ^{\tilde{n}}\sigma ^{\dagger }-\frac{1}{2}\gamma _{\uparrow }^{\tilde{n}%
n}\left\{ \sigma \sigma ^{\dagger },\rho ^{n}\right\} _{+}\right) .\ \ \ \  
\notag
\end{eqnarray}%
Thus, the non-diagonal terms lead to the transitions $|+,n\rangle \overset{%
\gamma _{\downarrow }^{n\pm 1,n}}{\rightarrow }|-,n\pm 1\rangle $ and
similarly $|-,n\rangle \overset{\gamma _{\uparrow }^{n\pm 1,n}}{\rightarrow }%
|+,n\pm 1\rangle $ [mechanisms (b) and (c) in Fig.~1].\ The rates satisfy%
\begin{equation}
\frac{\gamma _{\uparrow }^{n\tilde{n}}}{\gamma _{\downarrow }^{\tilde{n}n}}%
=e^{-\beta \lbrack E_{n}+\varepsilon _{+}^{(n)}-(E_{\tilde{n}}+\varepsilon
_{-}^{(\tilde{n})})]}.
\end{equation}%
In the continuous limit $(\beta \delta E\ll 1)$ the dynamics can be
approximated as%
\begin{eqnarray}
\frac{d\varrho _{x}}{dt} &\simeq &-i[H_{x},\varrho _{x}]+\mathcal{\tilde{L}}%
_{\mathrm{th}}^{(x)}[\varrho _{x}] \\
&&+\gamma \mathcal{L}_{\mathrm{FP}}^{(-)}[\sigma \varrho _{x}\sigma
^{\dagger }]+\gamma \mathcal{L}_{\mathrm{FP}}^{(+)}[\sigma ^{\dagger
}\varrho _{x}\sigma ],  \notag
\end{eqnarray}%
where $\mathcal{\tilde{L}}_{\mathrm{th}}^{(x)}$\ is the operator defined by
Eq.~(\ref{LthEne}) $(n=x/\delta x)$ but with renormalized ($x$-dependent)
rates. On the other hand, $\gamma $ remains as an arbitrary rate while $%
\mathcal{L}_{\mathrm{FP}}^{(\pm )}$ are set by Eq.~(\ref{Lfp}).

\section{Diffusive approximation under a detailed balance condition\label%
{QFPApprox}}

Here we sketch the derivation of the quantum Fokker-Planck-like dynamics
Eq.~(\ref{LatticeContinuas}). First, we rewrite the lattice dynamics~(\ref%
{Discrete}) as%
\begin{equation}
\frac{d\rho ^{n}}{dt}=\mathbb{L}_{n+1}[\rho ^{n+1}]+\mathbb{R}_{n-1}[\rho
^{n-1}]-(\mathbf{L}_{n}+\mathbf{R}_{n})[\rho ^{n}].  \label{Gen}
\end{equation}%
In this expression we did not consider diagonal couplings that can be worked
out straightforwardly under the replacement $n\rightarrow x.$ On the other
hand, for simplifying the notation we introduced the superoperators $\mathbb{%
L}_{n+1}$, $\mathbb{R}_{n-1},$ as well as $\mathbf{L}_{n}$ and $\mathbf{R}%
_{n}.$ Their action and structure can be read from Eq.~(\ref{Discrete}).

The diffusive approximation relies on a second order series expansion~\cite%
{kampen,budini}, $f_{n\pm 1}\simeq f(x)\pm (\partial /\partial x)f(x)\delta
x+(1/2)(\partial ^{2}/\partial ^{2}x)f(x)\delta x^{2}.$ To order cero in $%
\delta x,$ Eq. (\ref{Gen}) leads to%
\begin{equation}
\frac{d\varrho _{x}}{dt}\simeq (\mathbb{L}_{x}+\mathbb{R}_{x}-\mathbf{L}_{x}-%
\mathbf{R}_{x}\mathbb{)}[\varrho _{x}].
\end{equation}%
The second line of Eq.~(\ref{LatticeContinuas}) corresponds to this
contribution. In classical systems this cero order contribution vanishes
exactly due to the absence of any quantum superoperator, that is, when $%
\mathbb{L}_{x}=\mathbf{L}_{x}$ and $\mathbb{R}_{x}=\mathbf{R}_{x}.$

To first and second order in $\delta x,$ from Eq. (\ref{Gen}) it follows the
(extra) quantum Fokker-Planck equation%
\begin{equation}
\frac{d\varrho _{x}}{dt}\simeq -\delta x\frac{\partial }{\partial x}[(%
\mathbb{R}_{x}-\mathbb{L}_{x})\varrho _{x}]+\frac{\delta x^{2}}{2}\frac{%
\partial ^{2}}{\partial ^{2}x}(\mathbb{R}_{x}+\mathbb{L}_{x})\varrho _{x}.
\label{PreFP}
\end{equation}%
Notice that here all superoperators are of the same kind. Hence, their
action can be taken over $\varrho _{x}$ $[\Pi ^{s}\varrho _{x}\Pi ^{s}$ in
Eq.~(\ref{LatticeContinuas})], remaining only a standard Fokker-Planck
equation. Consequently, the following calculation steps are the same as in
classical systems (the expressions and notation are taken consistently with
this fact).

The detailed balance condition is introduced as follows. First, notice that
Eq.~(\ref{Discrete}) implies that $\mathbb{L}_{n+1}=\gamma ^{n,n+1}$ and $%
\mathbb{R}_{n-1}=\gamma ^{n,n-1}.$ The rate relations~(\ref{DetBalLattice})
are rewritten as $(\gamma ^{n,n+1}/\gamma ^{n+1,n})=e^{-\beta (\mathbb{E}%
_{n}-\mathbb{E}_{n+1})}$ and $(\gamma ^{n,n-1}/\gamma ^{n-1,n})=e^{-\beta (%
\mathbb{E}_{n}-\mathbb{E}_{n-1})},$ where, again, the energies $\{\mathbb{E}%
_{n}\}$ are introduced for simplifying the notation. These equalities imply
the (equivalent) relations 
\begin{equation}
\mathbb{L}_{n+1}=e^{-\beta (\mathbb{E}_{n}-\mathbb{E}_{n+1})}\mathbb{R}%
_{n},\ \ \ \ \ \ \mathbb{R}_{n-1}=e^{-\beta (\mathbb{E}_{n}-\mathbb{E}%
_{n-1})}\mathbb{L}_{n}.
\end{equation}%
Performing a Taylor series to first order in $\delta x$ it follows $%
[1+\delta x(\partial /\partial x)]\mathbb{L}_{x}\simeq \mathbb{R}%
_{x}[1+\beta \delta x(\partial /\partial x)\mathbb{E}_{x}]$ and similarly $%
[1-\delta x(\partial /\partial x)]\mathbb{R}_{x}\simeq \mathbb{L}%
_{x}[1-\beta \delta x(\partial /\partial x)\mathbb{E}_{x}].$ Consequently,%
\begin{equation}
(\mathbb{R}_{x}-\mathbb{L}_{x})\rightarrow -\gamma \beta \frac{\partial 
\mathbb{E}_{x}}{\partial x}\delta x/2,
\end{equation}%
where we have assumed that%
\begin{equation}
\mathbb{R}_{x}\rightarrow \gamma /2,\ \ \ \ \ \ \ \ \ \mathbb{L}%
_{x}\rightarrow \gamma /2.
\end{equation}%
In general, the rate $\gamma $ could depends on position, $\gamma
\rightarrow \gamma _{x}$ (when diffusion coefficient depends on position).
The last two relations, from Eq.~(\ref{PreFP}), imply that%
\begin{equation}
\frac{d\varrho _{x}}{dt}\simeq \gamma \frac{\delta x^{2}}{2}\left\{ \frac{%
\partial }{\partial x}[\beta \frac{\partial \mathbb{E}_{x}}{\partial x}%
\varrho _{x}]+\frac{\partial ^{2}}{\partial ^{2}x}\varrho _{x}\right\} ,
\end{equation}%
which in turn leads to the last contribution in Eq.~(\ref{LatticeContinuas}%
). Notice that in classical systems \cite{kampen} this equation has the
stationary solution $\varrho _{x}\approx \exp [-\beta \mathbb{E}_{x}].$

\end{document}